\documentclass[10pt,journal,compsoc]{IEEEtran}

\ifCLASSOPTIONcompsoc
  \usepackage[nocompress]{cite}
  \usepackage{graphicx}
  \usepackage{epstopdf}
  \usepackage{url}
\else
  \usepackage{cite}
  \usepackage{graphicx}
\fi
\ifCLASSINFOpdf
\else
\fi
\begin{document}

\title{Multi-Party Privacy-Preserving Record Linkage using Bloom Filters}

\author{Dinusha~Vatsalan and Peter~Christen 

\IEEEcompsocitemizethanks{\IEEEcompsocthanksitem Dinusha Vatsalan is a Postdoctoral
Research Fellow with the Research School of Computer Science,
The Australian National University, Canberra, Australia, ACT 2601.
E-mail: dinusha.vatsalan@anu.edu.au \protect\\

\IEEEcompsocthanksitem Peter Christen is a Professor with
the Research School of Computer Science,
The Australian National University, Canberra, Australia, ACT 2601.
E-mail: peter.christen@anu.edu.au 
}

\thanks{This paper is an extended version of the poster
paper published in proceedings of ACM Conference in Information
and Knowledge Management (CIKM) 2014~\cite{Vat14c}.}
}

\IEEEtitleabstractindextext{%
\begin{abstract}
Privacy-preserving record linkage (PPRL), the problem of identifying records that correspond to the same real-world entity across several data sources held by different parties without revealing any sensitive information about these records, is increasingly being required in many real-world application areas. Examples range from public health surveillance to crime and fraud detection, and national security. Various techniques have been developed to tackle the problem of PPRL, with the majority of them considering linking data from only two sources. However, in many real-world applications data from more than two sources need to be linked. In this paper we propose a viable solution for multi-party PPRL using two efficient privacy techniques: Bloom filter encoding and distributed secure summation. Our proposed protocol efficiently identifies matching sets of records held by all data sources that have a similarity above a certain minimum threshold. While being efficient, our protocol is also secure under the semi-honest adversary model in that no party can learn any sensitive information about any other parties' data, but all parties learn which of their records have a high similarity with records held by the other parties. We evaluate our protocol on a large real voter registration database showing the scalability, linkage quality, and privacy of our approach.
\end{abstract}

\begin{IEEEkeywords}
Record linkage, privacy, Bloom filters, secure summation, multi-party, approximate matching.
\end{IEEEkeywords}}

\maketitle

\IEEEdisplaynontitleabstractindextext

%
\IEEEpeerreviewmaketitle

\ifCLASSOPTIONcompsoc
\IEEEraisesectionheading{\section{Introduction}\label{sec:introduction}}
\else
\section{Introduction}
\label{sec:intro}
\fi

Many organizations collect and process datasets that contain many millions of records to analyze and mine interesting patterns and knowledge in order to support efficient and quality decision making~\cite{Fay96}. Analyzing and mining such large datasets often require data from multiple sources to be aggregated. Linking records from different data sources with the aim to improve data
quality or enrich data for further analysis is occurring in
an increasing number of application areas, such as in healthcare,
government services, crime and fraud detection, 
national security, and business
applications~\cite{Chr12,Elm07}.
The analysis of data linked across organizations can, for example,
facilitate the detection of an outbreak of an infectious disease early before it
spreads widely around a country or even worldwide, or enable the accurate
identification of fraud, crime, or terrorism suspects~\cite{Vat13}.
The health outbreak system described above requires
data from several organizations, such as
human health data, travel data, consumed drug data, and even animal health data~\cite{Clif02}.
The second above
example of fraud and crime detection
requires data from law enforcement agencies, Internet service
providers, businesses, as well as financial institutions~\cite{Phu12}.

Today, record linkage not only faces computational challenges due to
the increasing size of datasets, and quality challenges due to the
presence of real-world data errors, but also the challenge of preserving privacy and
confidentiality due to growing privacy concerns by the
public~\cite{Vat13,Clif02}. In the absence of unique entity identifiers in the databases
that are linked, personal identifying attributes (such as names,
addresses, gender, and dates of birth) are often used for the
linkage. Known as quasi-identifiers (QIDs)~\cite{Dal86}, values in
such attributes are in general sufficiently well correlated with
the corresponding real-world
entities to allow accurate linkage. Using such personal information
across different organizations,
however, often leads to privacy and confidentiality concerns. 

The privacy challenges posed in the record linkage process has led to the
development of techniques that facilitate `privacy-preserving record
linkage' (PPRL) ~\cite{Vat13}. PPRL tackles the problem of how to
identify records that refer to the same entity in different
databases across organizations
using the masked QIDs that are revealed. Generally,
the original QID values are transformed (masked) such that a specific
functional relationship exists between the original and the masked
values~\cite{Fie05}, and linkage is conducted using those masked QIDs 
without privacy and
confidentiality of the entities represented by the records
being compromised. 

A viable PPRL solution that can be used in real-world applications
needs to address all three challenges (or properties) of PPRL: scalability
(which is dependent on the computation and communication
complexity of a protocol),
linkage quality (dependent on data quality and the comparison 
functions and classifiers used), 
and privacy (dependent on the privacy techniques employed).
While there have been many different approaches proposed for
PPRL~\cite{Vat13}, most work in this research area thus far has concentrated on
linking records from only two sources (or parties). 
As the example
applications described above have shown, linking data from several
sources is however commonly required.

\begin{figure*}[t!]
\centering
\includegraphics[height=4.0cm]{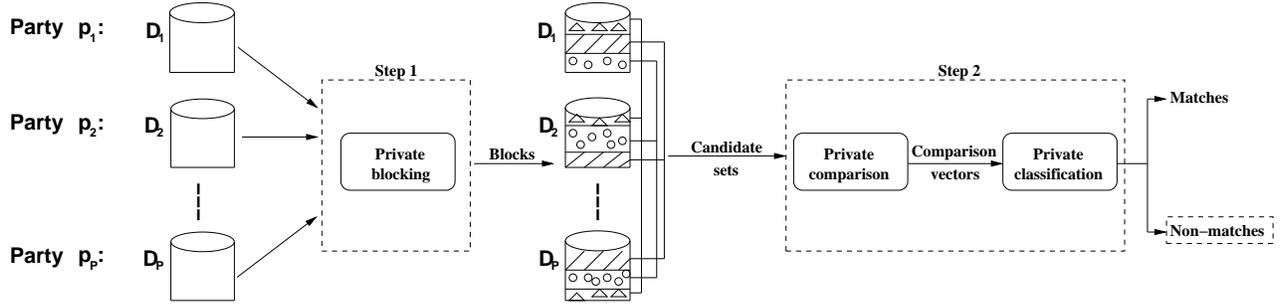}
\caption{A general privacy-preserving
record linkage pipeline for multiple ($P$) databases.
Databases are partitioned in step 1 to reduce the number of candidate sets such that only records in the same blocks or partitions are compared and classified in step 2. Everything represented with a dotted outline needs to be conducted such that privacy is preserved.
}
\label{fig:PPRL_multiple}
\end{figure*}

The pipeline of PPRL for multiple databases is shown in Fig.~\ref{fig:PPRL_multiple}.
PPRL on multiple databases introduces additional challenges with respect to scalability (complexity), linkage quality, and privacy~\cite{Chr14}. Complexity increases significantly with multiple parties in terms of both computational efforts and communication size. The number of all-to-all comparisons required between $P$ different databases ($\mathbf{D}_1,\mathbf{D}_2,\cdots,\mathbf{D}_P$) is equal to the product of the size of these databases (i.e.\ $|\mathbf{D}_1| \times |\mathbf{D}_2| \cdots \times |\mathbf{D}_P|$). As shown in Fig.~\ref{fig:PPRL_multiple}, the quadratic or exponential complexity (for linking two databases or multiple databases, respectively) has been addressed by using two-step algorithms where in the first step a private blocking approach is used in order to reduce the number of candidate record sets, that will then be compared and classified in the second step using private comparison and classification functions~\cite{Vat13}. However, in multi-party PPRL the total number of candidate record sets increases exponentially with the number of parties, and thus even using existing private blocking techniques would not sufficiently reduce the number of comparisons. Efficient and advanced blocking and filtering approaches for multi-party PPRL need to be used in order to reduce this potentially huge number of comparisons. Computations should also be distributed among the different parties to reduce the computational efforts at each individual party.

The risk of privacy breaches also increases with multiple parties due to possible collusion between a sub-set of parties with the aim to learn about another (sub-set of) party's private data. All computations should be distributed among the parties in such a way that each party can learn only a limited amount of information of other parties' data that cannot be used to infer the represented entities. Employing computationally expensive privacy techniques such as Secure Multi-party Computation (SMC)~\cite{Hal10,Lin09} provides strong privacy guarantees at the cost of increased computational and communication complexities with the increasing size of datasets and the increasing number of parties, making SMC solutions not scalable and practical in real applications.

With regard to quality, private comparison and classification on multiple databases is another challenging aspect due to the need of calculating the similarity of multiple values. How to efficiently calculate the similarity of more than two values using approximate comparison functions~\cite{Chr12} in PPRL is an important research question that needs to be addressed. Existing PPRL solutions for multiple parties only support exact matching (which classifies sets of records as matches if their masked QIDs are exactly the same and as non-matches if they differ)~\cite{Lai06,Qua98}, or they are applicable to QIDs of categorical data type only (while in most PPRL applications QIDs of string data type, such as names and addresses, are commonly required)~\cite{Kan08,Moh11}. 

Despite these challenges, PPRL on multiple databases is required in many real-world applications (as described above). 
A recent work by Ranbaduge et al.~\cite{Ran14} aimed to reduce the number of candidate record sets that need to be compared by using a multi-party private blocking approach. 
In this paper, we focus on the private comparison and classification step in the PPRL pipeline for multiple databases (step~2 in Fig.~\ref{fig:PPRL_multiple}) and propose a solution that performs efficient (distributed) and approximate matching of string values using computation and space efficient privacy techniques: Bloom filters and distributed secure summation. 
We also include a filtering approach into our protocol that can be used to considerably reduce the number of comparisons (in addition to a private blocking approach).

The main contributions of this paper are: (1) an efficient multi-party protocol for private comparison and classification in PPRL; (2) a filtering approach on candidate sets of records that are likely to correspond to non-matches; (3) an analysis of the protocol in terms of the three properties, scalability (complexity), linkage quality, and privacy; and (4) an empirical evaluation and comparison of our protocol with a baseline multi-party approach~\cite{Lai06} in terms of the three properties of PPRL using the large North Carolina Voter Registration (NCVR)~\cite{Chr13NC} datasets.

The remainder of the paper is structured as follows:
In the following section we review related work in
multi-party PPRL.
In Section~\ref{sec-protocol} we 
describe the steps of our multi-party protocol for
efficient and approximate private comparison and classification
in PPRL. 
We analyze our solution in terms of complexity, linkage quality, and privacy
in Section~\ref{sec-analysis}, and  
in Section~\ref{sec-experiment} we
conduct an empirical study on the large real NCVR datasets
to validate these analyses. Finally, we summarize and
discuss future research directions in
Section~\ref{sec-conclusion}.


\section{Related Work}
\label{sec-related}

Various techniques have been developed to address the PPRL research
problem~\cite{Vat13}, but few among these have considered PPRL on
multiple databases. An early approach to PPRL~\cite{Qua98} links
multiple databases by comparing the hash-encoded (using
one-way secure hash algorithms) QID values
from all data sources by using a third 
party. However, this approach only performs exact matching (i.e.\ a
single variation in a QID value results in a completely different
hash-encoded value). 

An SMC-based approach using an oblivious
transfer protocol was presented by O'Keefe et al.~\cite{Kee04} for PPRL on multiple
databases. 
The approach improves on 
the security and information leakage characteristics of several previous
protocols, including Agrawal et al.'s~\cite{Agr03} two-party secure
intersection and equi-join protocols that use commutative encryption
schemes. 
While provably secure, the approach only performs exact matching
of masked values (i.e.\ variations and errors in the QID
values are not considered). The approach is also computationally
expensive compared to perturbation-based privacy techniques~\cite{Vat13}. 

A multi-party approach based on the  
$k$-anonymity and secure equi-join privacy techniques
was introduced by Kantarcioglu et al.~\cite{Kan08}. 
The database owners individually $k$-anonymize their databases 
and send the anonymized databases to a third
party that constructs buckets
corresponding to each combination of $k$-anonymous values. For each
bucket, the third party performs a secure equi-join. 
This
approach is only applicable to categorical data.

Recently,
a multi-party PPRL approach for approximate matching of categorical
values based on $k$-anonymity was
proposed~\cite{Moh11}. 
The database owners find
the global winner candidate attribute with the best score that provides 
the least amount of 
information to the other parties according to some criteria. Then
they
perform a top-down specialization on that attribute for generalizing
the databases. The well-known C4.5
classifier is used to recursively block (generalize) and
classify the records in the databases.
Similar to~\cite{Kan08}, this approach is only applicable to
linking records using attributes that contain categorical data.

\begin{figure}[!t]
  \centering
  \scalebox{1.0}[1.0]{\includegraphics[width=0.5\textwidth]
                      {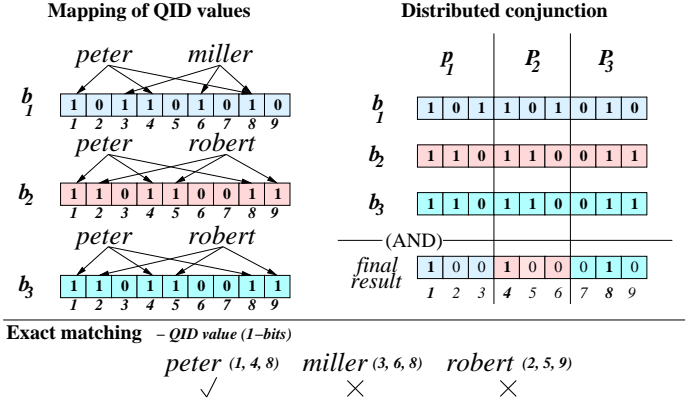}}
  \caption{Distributed exact matching of QIDs in multi-party PPRL ($P=3$)
           as proposed by Lai et al.~\cite{Lai06} (adopted from~\cite{Vat13}).
           Different colors represent the Bloom filters held by different
           parties and bit segments of Bloom filters processed by different parties.
           Bits 1, 4, and 8 are set to $1$ in the final result and therefore
           QID value `peter' is a matching value.}
           \label{fig-BF_MP_Lai}
\end{figure}

An efficient multi-party PPRL approach for exact matching 
using Bloom filters was introduced by Lai et al.~\cite{Lai06}. 
Fig.~\ref{fig-BF_MP_Lai} illustrates this approach
for $P=3$ example databases.  
The database values (QIDs) are first converted into one Bloom filter
$b_i, 1 \le i \le P$ per party.
Each party then partitions its Bloom filter into segments
according to the number of parties involved in the linkage, and sends
these segments to the corresponding other parties. The segments received by a party
are combined using a conjunction (logical AND) operation. The
final conjuncted Bloom filter segments are then exchanged among
the parties. Each party compares its Bloom filter of each QID value
with the
final result, and if the membership test of a QID value 
is successful then it
is considered to be a match. 
Though the cost of this approach is low
since the computation is completely distributed among the parties
and the creation and processing of Bloom filters are very fast
(linear complexity in the size of the database), the
approach can only perform exact matching. 

Since existing private comparison and classification solutions for
multi-party PPRL either (1) support \emph{exact matching} only (which is not applicable
in most real-world applications due to the common occurrences of data errors and variations~\cite{Her98}),
(2) employ \emph{expensive privacy} techniques such as SMC, or (3) they are only applicable to
\emph{categorical} data, we aim to overcome these three problems by
proposing a multi-party \emph{approximate string} matching protocol for PPRL 
using \emph{efficient privacy} techniques.
As we describe in the
next section, we use
Lai et al.'s~\cite{Lai06} multi-party Bloom filter based 
exact matching approach (described above) as one 
building block for our approximate matching solution.


\section{Multi-Party Linkage Protocol}
\label{sec-protocol}

We now describe our approach to efficiently and approximately link
databases from three or more parties. We use the following notation:
$P$ is the number of parties involved in our protocol, where each
party $p_i$ holds a database $\mathbf{D}_i$ containing sensitive or
confidential identifying information. Database $\mathbf{D}_i$ contains
$N_i=|\mathbf{D}_i|$ records. We assume a set of QID attributes $A$, which will
be used for the linkage, is common to all these databases. 
Our protocol will calculate the similarity between sets of records
using the values in $A$. In the following sub-section we
describe the building blocks of our protocol, then 
in Section~\ref{sec-steps} we explain the
steps of our protocol in detail, and 
in
Section~\ref{subsec:filter} 
we propose
a filtering approach to 
improve the efficiency of our protocol.

\subsection{Protocol Building Blocks}
\label{sec-blocks}

\textbf{1. Bloom filter encoding:}~
A Bloom filter $b_i$ is a bit array data structure of length $l$ bits
where all bits are initially set to $0$. $k$ independent hash
functions, $h_1,h_2, \ldots, h_k$, each with range $1, \ldots l$, are
used to map each of the elements in a set $S$ into the Bloom filter by
setting $k$ corresponding bit positions to $1$. Bloom filters are one
efficient perturbation-based privacy technique that has successfully
been used in several PPRL
solutions~\cite{Sch09,Dur13,Vat12}. 

Schnell et al.~\cite{Sch09} were the first to propose a method for
approximate matching in PPRL of two databases using Bloom filters. In
their work, as in our protocol, the character $q$-grams (sub-strings of length
$q$) of QID values in $A$ of each record in the databases to be
linked are hash-mapped into a Bloom filter using $k$ independent hash
functions. These Bloom filters are then sent to a third party that
calculates the Dice coefficient~\cite{Chr12} similarity of pairs of
Bloom filters.

Bloom filters can be susceptible to frequency attacks~\cite{Kuz11}
depending on the values of the parameters $k$, $l$, and $q$.
Hence, these Bloom filter parameters need to be set carefully as the values provide
a trade-off between privacy and linkage quality (as will be discussed
in Section~\ref{sec-analysis}).
Several Bloom filter encoding methods~\cite{Sch11,Dur13,Vat14} 
have been proposed
to improve privacy by reducing the risk of such frequency attacks
while not compromising the linkage quality. 

Schnell et al.~\cite{Sch11} proposed to hash-map several QID attribute
values of a record 
into one Bloom filter, known as Cryptographic Long term Key (CLK) encoding.
Durham et al.~\cite{Dur13} investigated composite Bloom filters
(record-level Bloom filters) in detail
by first hash-mapping different attributes into attribute-level
Bloom filters of different lengths 
(depending on the weights~\cite{Fel69} of QID attributes that calculate
the discriminatory power in resolving identity using a statistical approach) 
and then combining
these attribute-level Bloom filters into one record-level Bloom filter
(known as RBF) by sampling
bits from each attribute-level Bloom filter. Vatsalan et al.~\cite{Vat14}
recently introduced a hybrid method of CLK and RBF (known as CLKRBF) where
the Bloom filter length is kept to be the same as in CLK while using different number of
hash functions to map different attributes into the Bloom filter based on their weights
as used in RBF. 
\vspace*{4.5mm}

\begin{figure}[!t]
  \centering
  \scalebox{1.0}[1.0]{\includegraphics[width=0.4\textwidth]
                      {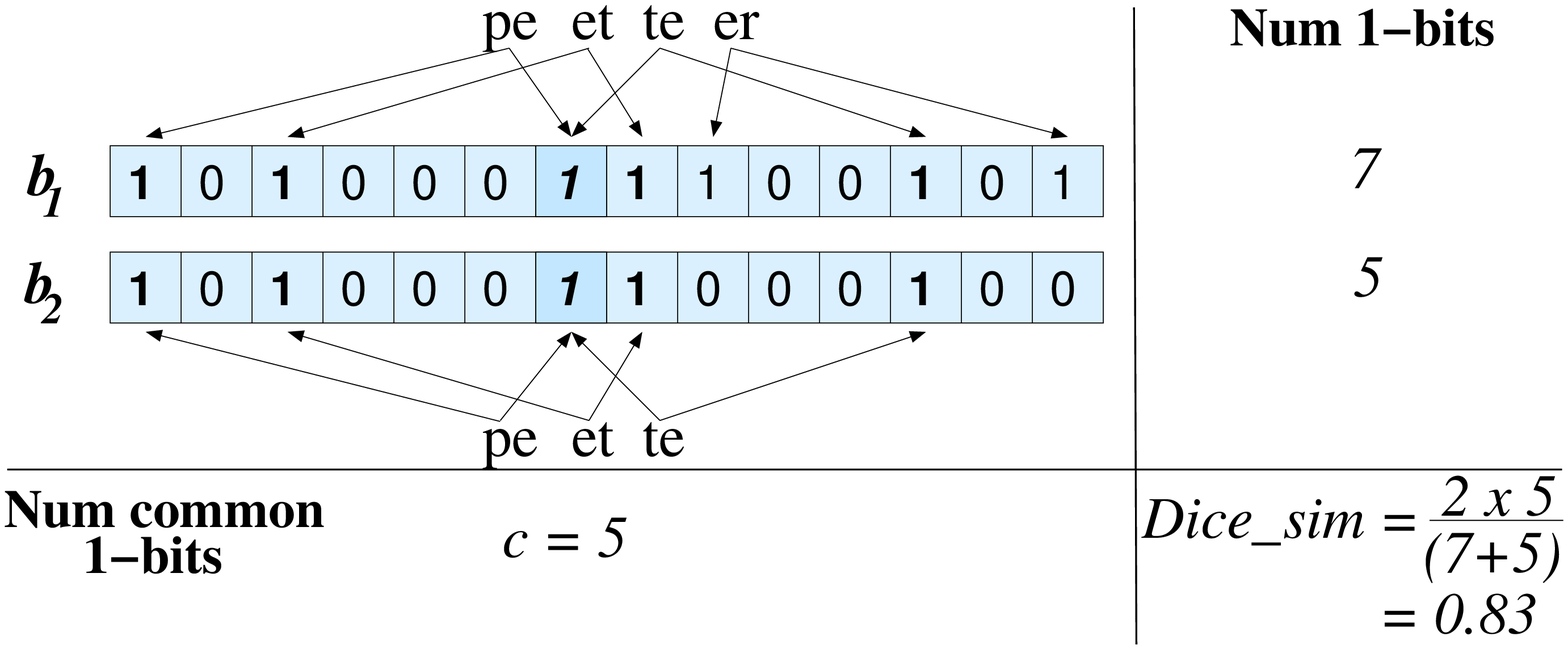}}
  \caption{Mapping of strings (QIDs) into Bloom filters and
           calculating their Dice coefficient similarity (adapted from~\cite{Vat13}).}
           \label{fig-BF}
\end{figure}

\noindent
\textbf{2. Dice coefficient:}~
Any set-based similarity function can be used to calculate the
similarity of pairs or sets of Bloom filters. The Dice coefficient has
previously been used for matching of Bloom filters in PPRL
since it is insensitive to
many matching zeros in long Bloom filters~\cite{Sch09}. 
The Dice coefficient similarity of two
Bloom filters ($b_1,b_2$) is calculated as~\cite{Chr12}:
\begin{eqnarray}
\label{eq:Dice_coefficient}
Dice\_sim(b_1,b_2) &=& \frac{2 \times c}{x_1 + x_2} 
\end{eqnarray}
where $c$ is the number of common bit positions that are set to $1$ in
both Bloom filters $b_1$ and $b_2$ (common $1$-bits), $x_1$ is the
number of bit positions that are set to $1$ in $b_1$, and $x_2$ is the
number of bit positions that are set to $1$ in $b_2$.
For example, mapping the
bigrams ($q=2$) of two string values `peter' and `pete' into
$l=14$ bits long Bloom filters using $k=2$ hash functions and 
calculating the Dice coefficient similarity of these two Bloom
filters are illustrated in
Fig.~\ref{fig-BF}. 

We define 
the Dice coefficient similarity of $P$ ($P \ge 2$) Bloom filters ($b_1, \cdots,
b_P$) as:
\begin{eqnarray}
\label{eq:Dice_coefficient}
Dice\_sim (b_1, \cdots, b_P) &=& \frac{P \times c}{\sum_{i=1}^{P} x_i} 
\end{eqnarray}
where $c$ is the number of common bit positions that are set to $1$ in
all $P$ Bloom filters (common $1$-bits), and $x_i$ is the number
of bit positions set to $1$ in $b_i$ ($1$-bits), $1 \le i \le P$. 
\vspace*{4.5mm}

\begin{figure}[!t]
  \centering
  \scalebox{1.0}[1.0]{\includegraphics[width=0.45\textwidth]
                      {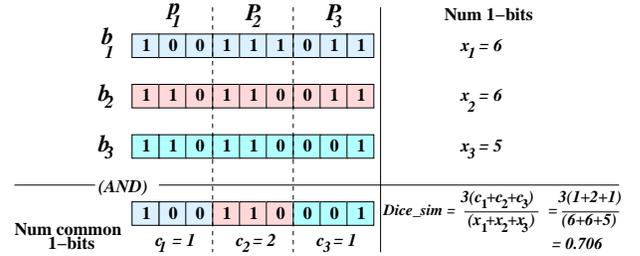}}
  \caption{Dice coefficient similarity calculation of three Bloom
           filters (BFs) across three parties ($P=3$). 
           Rows illustrate the BFs, $b_i$, generated by the three
           parties, $p_i$, with $1 \le i \le 3$, 
           while columns show which party holds which BF
           segments. Different colors represent the BFs held by different
           parties and the bit segments of BFs processed by different parties
           (adapted from~\cite{Vat14c}).}
           \label{fig-label_multi_party_BF}
\end{figure}

\noindent
\textbf{3. Multi-party Bloom filter matching:}~
In our protocol the calculation of the number of common $1$-bits ($c$)
is distributed among the parties, such that $c = \sum_{i=1}^{P} c_i$. 
\begin{eqnarray}
\label{eq:Dice_coefficient_mp}
Dice\_sim (b_1, \cdots, b_P) &=& \frac{P \times \sum_{i=1}^P c_i}{\sum_{i=1}^{P} x_i} 
\end{eqnarray}
Following Lai et al.'s approach~\cite{Lai06}, Bloom filters are split
into $P$ segments and each party sends its segments to the
corresponding other parties. Each party then individually calculates
the number of common $1$-bits $c_i$ in its respective segment of
the Bloom filters it receives from the other parties for all
sets of records. As an example, the distributed Dice coefficient
calculation of a set of three Bloom filters from three parties is shown
in Fig.~\ref{fig-label_multi_party_BF}. 
\vspace*{4.5mm}

\begin{figure}[!h]
  \centering
  \scalebox{0.7}[0.7]{\includegraphics[width=0.45\textwidth]
                      {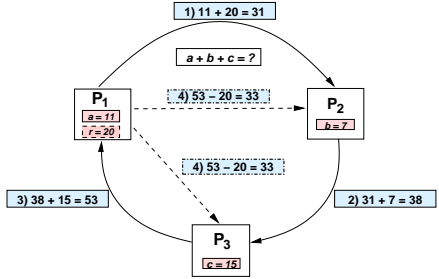}}
  \caption{Secure summation of a set of three private values ($a=11$, $b=7$, and $c=15$)
  using a random value $r=20$
in a ring-based communication pattern between three different parties
($p_1$, $p_2$, and $p_3$, respectively). Four communication phases
are involved in the secure summation of three values.}
   \label{fig-label_secure_sum}
\end{figure}

\noindent
\textbf{4. Secure summation:}~
Once each of the $P$ parties has calculated its $c_i$ and $x_i$ values
for each set of Bloom filters, the summations of values
$c = \sum_{i=1}^{P}
c_i$ and $x = \sum_{i=1}^{P} x_i$
need to be calculated  in a secure way in order to
calculate the Dice coefficient similarity of the set of Bloom filters. 
A secure summation protocol~\cite{Kar04}, which has been used as an efficient tool for
privacy-preserving data mining~\cite{Clif02}, 
can be efficiently employed for this
purpose.
This
protocol uses a random number $r$ (any integer number) to hide the
actual sensitive values $c_i$ and $x_i$, and employs a ring-based
communication pattern over all parties which allows each party to
learn the final values $c$ and $x$, but no party will learn the
individual values (i.e. $c_i$ and $x_i$) of the other parties. A simple example illustrating
the secure summation protocol is shown in Fig.~\ref{fig-label_secure_sum}.

\begin{figure*}[!t]
  \centering
  \scalebox{2.0}[2.3]{\includegraphics[width=0.5\textwidth]
                      {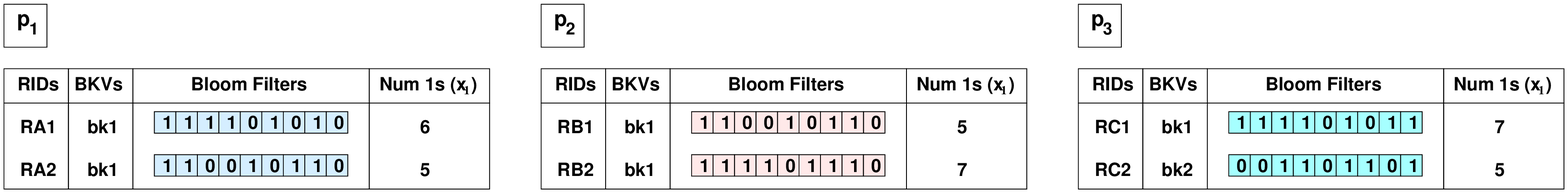}}
  \caption{Example Bloom filters held by three parties $p_1$, $p_2$,
            and $p_3$, and the number of $1$-bits ($x_1$, $x_2$, and $x_3$, respectively)
           in each of their Bloom filters along with the RIDs 
           stored in $\mathbf{D}_1$, $\mathbf{D}_2$, and $\mathbf{D}_3$ respectively,
           and the BKVs, 
            used to illustrate the steps of the protocol described in Section~\ref{sec-steps}.
            }
           \label{fig-label_mpbf1}
\end{figure*}

\begin{figure*}[!t]
  \centering
  \scalebox{1.4}[1.3]{\includegraphics[width=0.5\textwidth]
                      {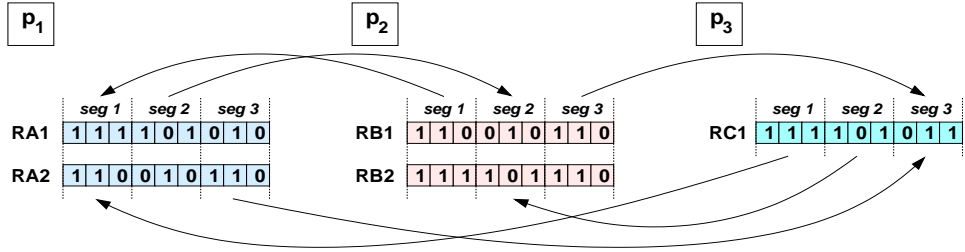}}
  \caption{The splitting of Bloom filters (in the block `bk1') into $P$ segments and exchanging
           the corresponding Bloom filter segments among the parties such
            that each party $p_i$ receives the $i^{th}$ 
            Bloom filter segments from all other parties.
           This figure illustrates Step~4 of the protocol.}
           \label{fig-label_mpbf2}
\end{figure*}

\subsection{Protocol Steps}
\label{sec-steps}

In this section we describe in detail the steps of our protocol
to approximately and privately link databases from
$P$ $(\ge 3)$ sources/parties. 
We illustrate the steps using 
three example datasets held by three parties, 
as shown in Fig.~\ref{fig-label_mpbf1}. In Fig.~\ref{fig-label_mpbf2} 
to Fig.~\ref{fig-label_mpbf4} we illustrate the steps of our protocol for 
the three example datasets. 

\begin{itemize}

\item \textbf{Step 1:}~
The parties agree upon the following parameter values:
      the Bloom filter length $l$ such that $l~mod~P = 0$
      to allow splitting of Bloom filters into segments of same size;
      the $k$ hashing 
      functions $h_1, \ldots, h_k$ to be used; the length (in characters) of 
      grams $q$; a minimum Dice similarity threshold value, $s_t$,
      above which a set of records is classified as a match;
      a private blocking function $block(\cdot)$; 
      the blocking keys~\cite{Chr12} $B$ used for blocking; 
      and a set of QID attributes $A$ used for the linkage.

      The setting of Bloom filter parameters and the encoding method
      is crucial to determine the privacy of our protocol.
      We propose to perform a simulation attack~\cite{Vat12} by the database
      owners on their own sets of Bloom filters in terms of the
      sensitivity of each bit in the Bloom filters
      before agreeing on the parameter setting, as will be
      discussed in detail in Section~\ref{subsec_privacy_analsis}.
      
\item \textbf{Step 2:}~
Each party $p_i$ ($1 \le i \le P$) individually 
applies a private blocking
function~\cite{Vat13} $block(\cdot)$ (step~1 in Fig.~\ref{fig:PPRL_multiple})
to reduce the number of candidate sets of
records (from $\prod_i^P N_i$). It is important to use a 
blocking function as the total number of sets of records from
$P$ databases quickly becomes prohibitive even for moderate
$P$ or $N$. 
$block(\cdot)$ groups records according to the
blocking key values (BKVs)~\cite{Chr11} and only 
records with the same BKV 
(i.e. records in the same block)
from different parties are then 
compared and classified (step~2 in Fig.~\ref{fig:PPRL_multiple})
using our protocol.

In the running example
we consider the five records with record identifiers (RIDs) 
$RA1$, $RA2$, $RB1$, $RB2$, and $RC1$ 
in the three databases which we assume
are blocked into the same block (i.e.\ have the same BKV - `bk1',
while record $RC2$ having a different BKV - `bk2'), so that
there
exist the following four candidate sets of (three) records from the
three parties (excluding sets of records from the same party):
$(RA1,RB1,RC1)$, $(RA1,RB2,RC1)$, $(RA2,RB1,RC1)$, and $(RA2,RB2,RC1)$
for comparison and classification.

\item \textbf{Step 3:}~
Each party $p_i$
hash-maps the $q$-gram values of $A$ of each of
its $N_i$ records in their respective databases $\mathbf{D}_i$
into $N_i$ Bloom filters of length $l$ using
the hash functions $h_1, \ldots, h_k$. It is crucial
to set the Bloom filter related parameters in an optimal
way that balances all three properties of PPRL 
(complexity, quality, and privacy). We further discuss the parameter
setting for Bloom filters used in our protocol
in Section~\ref{sec-analysis}.
For all records and their Bloom filters, each party $p_i$
calculates the total number of $1$-bits in the Bloom filters
($x_i$) and stores these values along with  
RIDs and BKVs, as shown 
in Fig.~\ref{fig-label_mpbf1}.

\item \textbf{Step 4:}~
Each party $p_i$
segments its Bloom filters into $P$ equal sized segments
of length $l/P$ bits and
sends the ${j}^{th}$ segment of each of its
Bloom filters along with the (encrypted) RIDs and
BKVs to party $p_j$, with $1 \le j \le
P$ and $j \neq i$. This step is illustrated 
for the three example datasets in Fig.~\ref{fig-label_mpbf2}.

\begin{figure*}[!t]
  \centering
  \scalebox{2.0}[1.3]{\includegraphics[width=0.5\textwidth]
                      {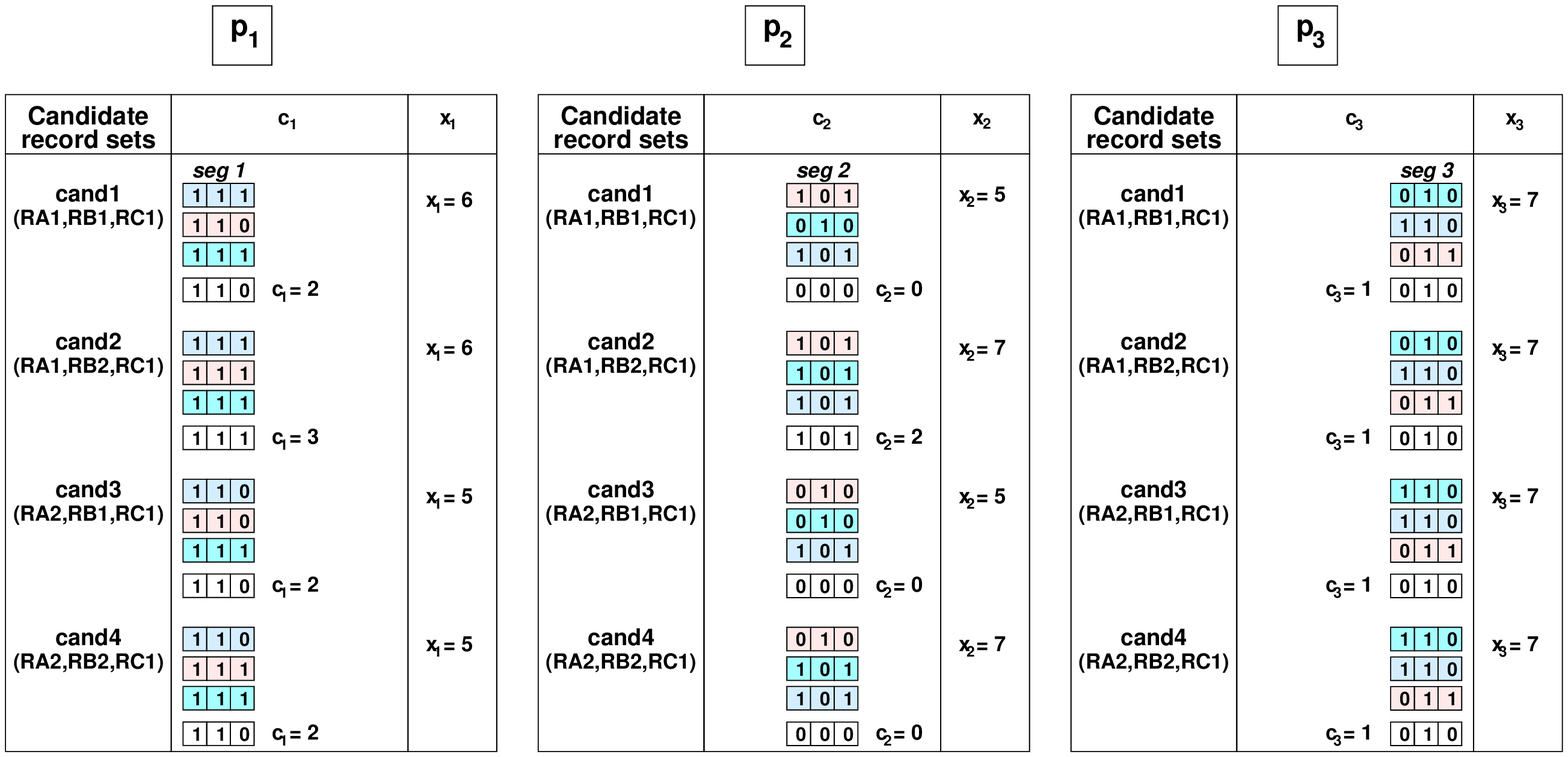}}
  \caption{The calculation of values for $c_i$ and $x_i$
           individually by each party $p_i$
           for all the candidate sets of records from all three parties.
           Different colors represent the Bloom filter segments
           received from different parties.
           This figure illustrates Step~5 of the protocol.
           }
           \label{fig-label_mpbf3}
\end{figure*}

\begin{table}[!t]
  \centering
  \small\addtolength{\tabcolsep}{-4pt}
  \label{algo_dist_calc}
  \begin{tabular}{ll} \hline
    ~ \\[-2mm]
    \multicolumn{2}{l}{\textbf{Algo.~1:} Distributed calculation of common $1$-bits ($c_i$) by $p_i$.}
      \\[1mm] \hline
    ~ \\[-3mm]
    \multicolumn{2}{l}{\textbf{Input:}} \\
    \multicolumn{2}{l}{- $my\_segs$: 
      List of record IDs, their BKVs, $i^{th}$ Bloom filter} \\
    ~ & {~ segments, and total number of $1$-bits $(x_i)$ in the Bloom} \\
    ~ & {~ filters held by party $p_i$} \\
    \multicolumn{2}{l}{- $other\_segs$: 
      Lists of record IDs, their BKVs, and $i^{th}$ Bloom} \\
    ~ & {~ filter segments of other parties $p_j$, $1 \le j \le P$ and $j \neq i$} \\
    \multicolumn{2}{l}{- $common\_blocks$: List of BKVs common in all $P$ databases} \\
    \multicolumn{2}{l}{\textbf{Output:}} \\
    \multicolumn{2}{l}{- $\mathbf{C}_i$: Candidate record sets with their $c_i$ and $x_i$ values} \\[1mm]
    1:  & $\mathbf{C}_i = [\,]$ \\ 
    2:  & \textbf{for} $ bkv \in common\_blocks$ \textbf{do}: \\
    3:  & ~~ $candidate\_sets = my\_segs.get\_RIDs(bkv)$ \\
    4:  & ~~ \textbf{for} $1 \le j \le P$ and $j \neq i$ \textbf{do}: \\
    5:  & ~~ ~~ $p_j\_records = other\_segs.get\_RIDs(bkv)$ \\ 
    6:  & ~~ ~~ \textbf{for} $rec\_comb \in candidate\_sets$ \textbf{do}: \\
    7:  & ~~ ~~ ~~ \textbf{for} $rec \in p_j\_records$ \textbf{do}: \\
    8:  & ~~ ~~ ~~ ~~ $rec\_comb.append(rec)$ \\
    9:  & ~~ \textbf{for} $cand \in candidate\_sets$ \textbf{do}: \\ 
    10: & ~~ ~~ $x_i = my\_segs.get\_x_i(cand[i])$ \\
    11: & ~~ ~~ $bf\_seg = my\_segs.get\_seg(cand[i]) $ \\
    12: & ~~ ~~ \textbf{for} $1 \le j \le P$ and $j \neq i$ \textbf{do}: \\
    13: & ~~ ~~ ~~ $bf\_seg$ \& $= other\_segs.get\_seg(cand[j])$ \\
    14: & ~~ ~~ $c_i = bf\_seg.count\_1bits() $ \\
    15: & ~~ ~~ $\mathbf{C}_i.append([cand,c_i,x_i])$ \\[1mm]
      \hline
  \end{tabular}
\end{table}

\item \textbf{Step 5:}~
Each party $p_i$
receives the $i^{th}$ segment of Bloom filters from all other 
parties $p_j$, with $1 \le i,j \le P$ and $j \neq i$.
For each set of Bloom filters ($b_1, b_2, \cdots , b_P$)
of the records from all parties
that are in the same block, party $p_i$ applies 
a logical conjunction (AND) on the Bloom filter
segments ($b_1^i \wedge b_2^i \wedge \cdots \wedge b_P^i$).
This results in the common bit pattern for segment $i$ from all parties
which allows party $p_i$
to calculate the number of common $1$-bits ($c_i$)
in the $i^{th}$ segment.
Fig.~\ref{fig-label_mpbf3} illustrates this distributed calculation 
of $c_i$ values for the running example candidate sets (in block `bk1').

The distributed common $1$-bits calculation ($c_i$, $1 \le i \le P$)
is described in Algo.~1
for one party $p_i$ (this algorithm is
executed by each party individually).
The party first generates the candidate sets of records
$candidate\_sets$ in lines~1-8.
For each candidate set $cand$ in $candidate\_sets$ 
the $i^{th}$ segments from all parties are conjuncted ($ \& $) 
in lines~11-13 to generate the bit
pattern of that segment that contains only
the common $1$-bits.
The number of common $1$-bits in the $i^{th}$
segment ($c_i$) is calculated in line~14 
for each candidate set by using a \emph{count\_1bits()} function 
and stored in $\mathbf{C}_i$ along with the number of $1$-bits in the
full Bloom filter of party $p_i$'s record in the set
($x_i$, which is calculated in line~10), and the
list of all RIDs ($cand$) in the set. 

\begin{figure*}[!t]
  \centering
  \scalebox{2.0}[1.3]{\includegraphics[width=0.5\textwidth]
                      {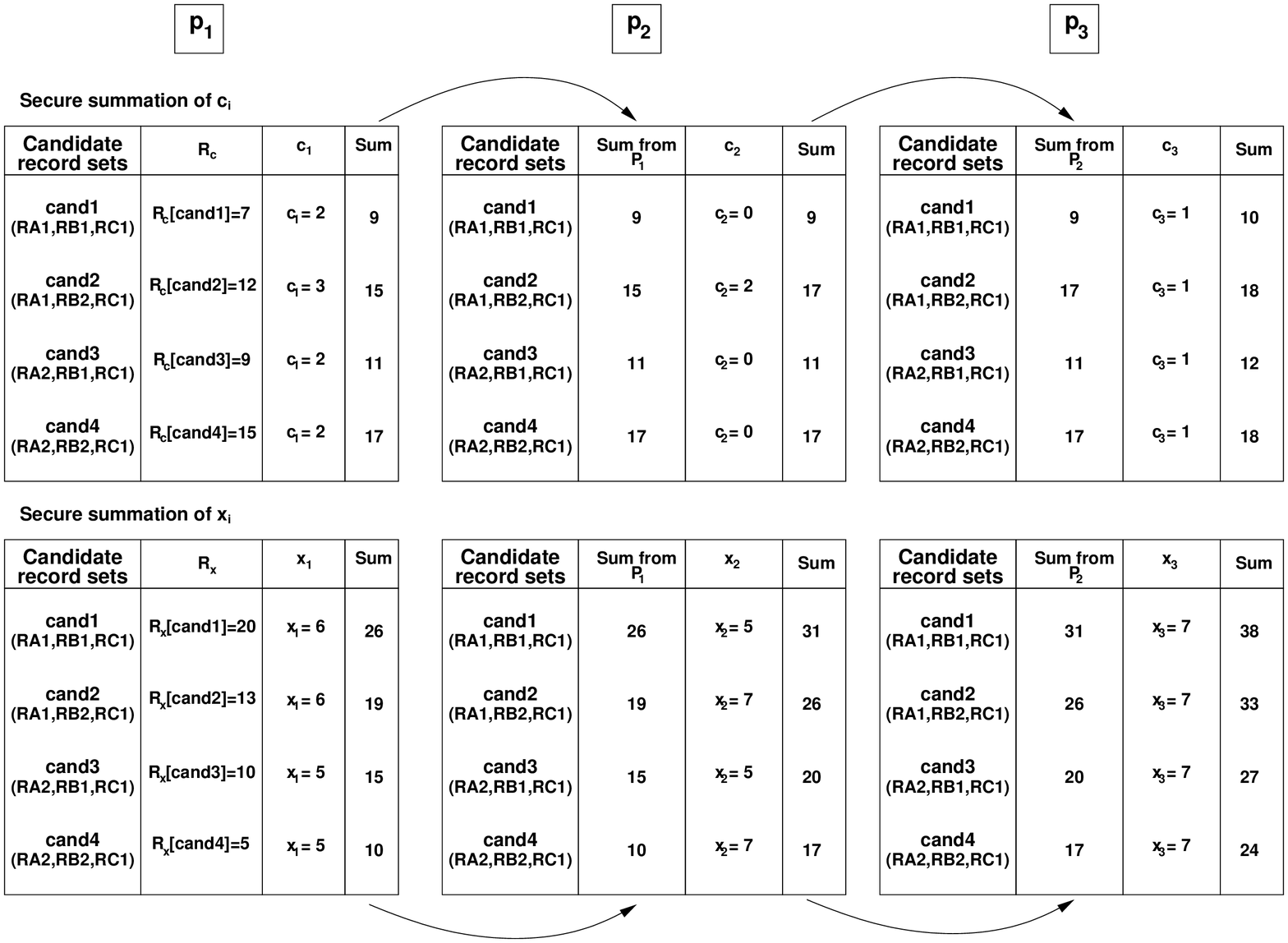}}
  \caption{The secure summation of the $c_i$ and $x_i$ values to calculate
           $c=\sum_{i=1}^P c_i$ and $x=\sum_{i=1}^P x_i$, respectively,
           for each candidate set of records,
           in order to calculate the Dice coefficient similarity of
           those record sets. 
           The lists of random values 
           used by party $p_1$ for the secure summation of the $c_i$
           and $x_i$ values are $\mathbf{R}_c=[7,12,9,15]$ 
           and $\mathbf{R}_x=[20,13,10,5]$,
           respectively.
           This figure illustrates Step~6 of the protocol.
           }
           \label{fig-label_mpbf4}
\end{figure*}

\begin{table}[!t]
  \label{algo_sec_sim_calc}
  \centering
  \small\addtolength{\tabcolsep}{-4pt}
  \begin{tabular}{ll} \hline
    ~ \\[-2mm]
    \multicolumn{2}{l}{\textbf{Algo.~2:} Secure summation of $c_i$ and $x_i$ values.}
      \\[1mm] \hline
    ~ \\[-3mm]
    \multicolumn{2}{l}{\textbf{Input:}} \\
    \multicolumn{2}{l}{- $\mathbf{C}_i$: 
      Candidate record sets of party $p_i$ with $c_i$ and $x_i$} \\
      ~ & ~ {values, $1 \le i \le P$} \\
      \multicolumn{2}{l}{- $\mathbf{R}_c$ and $\mathbf{R}_x$: Lists of random values used by party $p_1$ for} \\
      ~ & ~ {secure summation of $c_i$ and $x_i$ values, respectively} \\
    \multicolumn{2}{l}{\textbf{Output:}} \\
    \multicolumn{2}{l}{- $\mathbf{C'}$: Candidate record sets with summed values of $c_i$ and $x_i$} \\[1mm]
    1:  & $\mathbf{C'} = [\,]$ \\ 
    2:  & \textbf{for} $1 \le i \le P$ \textbf{do}: \\
    3:  & ~~ \textbf{if} $i == 1$ \textbf{then}: \\
    4:  & ~~ ~~ \textbf{for} $cand \in \mathbf{C}_1$ \textbf{do}: \\
    5:  & ~~ ~~ ~~ $c_i = \mathbf{R}_c[cand] + \mathbf{C}_1.get\_c_i(cand)$ \\ 
    6:  & ~~ ~~ ~~ $x_i = \mathbf{R}_x[cand] + \mathbf{C}_1.get\_x_i(cand)$ \\ 
    7:  & ~~ ~~ ~~ $\mathbf{C'}.append([cand,c_i,x_i])$ \\ 
    8:  & ~~ ~~ $\mathbf{C'}.send\_to(p_2)$ \\
    9:  & ~~ \textbf{else}: \\
    10:  & ~~ ~~ $\mathbf{C'}.receive\_from(p_{i-1})$ \\
    11:  & ~~ ~~ \textbf{for} $cand \in \mathbf{C'}$ \textbf{do}: \\
    12:  & ~~ ~~ ~~ $\mathbf{C'}.get\_c_i(cand)$ +$= \mathbf{C}_i.get\_c_i(cand)$ \\ 
    13:  & ~~ ~~ ~~ $\mathbf{C'}.get\_x_i(cand)$ +$= \mathbf{C}_i.get\_x_i(cand)$ \\ 
    14:  & ~~ ~~ \textbf{if} $i \neq P$ \textbf{then}: \\ 
    15:  & ~~ ~~ ~~ $\mathbf{C'}.send\_to(p_{i+1})$ \\ [1mm]
    16:  & ~~ ~~ \textbf{else}:  \\ 
    17:  & ~~ ~~ ~~ $\mathbf{C'}.send\_to(p_{1})$ \\ [1mm]
      \hline
  \end{tabular}
\end{table}

\item \textbf{Step 6:}~
Once the common $1$-bits in each segment $c_i$ are calculated by each
respective party $p_i$, a ring-based communication pattern is
used among the parties to securely calculate the summation of the
$c_i$ and $x_i$ values, $c=\sum_{i=1}^P c_i$ and $x=\sum_{i=1}^P x_i$, respectively, 
for each candidate set using the secure summation protocol,
as illustrated in Fig.~\ref{fig-label_mpbf4} for the running example.

\begin{table}[!t]
  \label{algo_classify}
  \centering
  \small\addtolength{\tabcolsep}{-4pt}
  \begin{tabular}{ll} \hline
    ~ \\[-2mm]
    \multicolumn{2}{l}{\textbf{Algo.~3:} Similarity calculation of record sets (by $p_1$).}
      \\[1mm] \hline
    ~ \\[-3mm]
    \multicolumn{2}{l}{\textbf{Input:}} \\
    \multicolumn{2}{l}{- $\mathbf{C'}$: 
      Candidate record sets with summed values of $c_i$ and $x_i$} \\
     ~ & ~ {from party $p_{P}$} \\
     \multicolumn{2}{l}{- $\mathbf{R}_c$ and $\mathbf{R}_x$: Lists of random values used by party $p_1$ for} \\
      ~ & ~ {secure summation of $c_i$ and $x_i$ values, respectively} \\
      \multicolumn{2}{l}{- $s_t$: Minimum similarity threshold to classify record sets} \\
    \multicolumn{2}{l}{\textbf{Output:}} \\
    \multicolumn{2}{l}{- $\mathbf{M}$: List of matching record sets} \\[1mm]
    1:  & $\mathbf{M} = [\,]$ \\ 
    2:  & $\mathbf{C'}.receive\_from(p_P)$ \\
    3:  & \textbf{for} $cand \in \mathbf{C'}$ \textbf{do}: \\
    4:  & ~~ $sum\_c_i = \mathbf{C'}.get\_c_i(cand) - \mathbf{R}_c[cand]$ \\ 
    5:  & ~~ $sum\_x_i = \mathbf{C'}.get\_x_i(cand) - \mathbf{R}_x[cand]$ \\ 
    6:  & ~~ $Dice\_sim(cand) = \frac{P \times sum\_c_i}{sum\_x_i}$ \\ 
    7:  & ~~ \textbf{if} $Dice\_sim(cand) \ge s_t$ \textbf{then}: \\ 
    8:  & ~~ ~~ $\mathbf{M}.append([cand,Dice\_sim(cand)])$ \\
    9.  & \textbf{for} $2 \le j \le P$ \textbf{do}: \\
    10:  & ~~ $\mathbf{M}.send\_to(p_j)$ \\ [1mm]
      \hline
  \end{tabular}
\end{table} 

Algo.~2 provides an overview of       
the secure summation of common and total $1$-bits
($\sum_{i=1}^P c_i$ and $\sum_{i=1}^P x_i$) for each candidate set.
The party that initiated the communication
(we assume the first party, $p_1$)
adds two random values $\mathbf{R}_c[cand]$ and $\mathbf{R}_x[cand]$
with its values for $c_i$ and $x_i$ ($i=1$), respectively, for
each candidate set $cand$, and
sends the summed values $\mathbf{R}_c[cand]+c_i$ and 
$\mathbf{R}_x[cand]+x_i$ to party $p_{i+1}$ (i.e. $p_2$)
in lines~3-8.
Party $p_{i}$, $1 < i \le P$ receives the summed values from $p_{i-1}$ and
adds its values for $c_{i}$ and $x_{i}$ for each candidate set
and sends the summed values to the next party $p_{i+1}$.
This process is repeated until the last party (i.e.\ $p_P$)
sums its $c_P$ and $x_P$ values with the received summed values
$\mathbf{R}_c[cand] + \sum_{i=1}^{P-1} c_i$ and $\mathbf{R}_x[cand] + \sum_{i=1}^{P-1} x_i$ from
party $p_{P-1}$, respectively, and sends the final summed values to $p_1$
(as explained in lines~9-17 in Algo.~2).

\item \textbf{Step 7:}~
Finally, the first party, $p_1$, that initiated the communication
subtracts $\mathbf{R}_c[cand]$ and $\mathbf{R}_x[cand]$ from the received final summed values
$\mathbf{R}_c[cand] + \sum_{i=1}^{P} c_i$ and $\mathbf{R}_x[cand] + \sum_{i=1}^{P} x_i$, respectively, 
for each candidate set $cand$ from
the last party $p_P$. This is outlined in Algo.~3
(lines~2-5) and illustrated in Fig.~\ref{fig-label_mpbf5} for the running
example. As shown in lines~6-8, $p_1$ then 
calculates the Dice coefficient similarity of each set of Bloom
      filters using $\sum_{i=1}^P c_i$ and $\sum_{i=1}^P x_i$ following
      Equation~\ref{eq:Dice_coefficient_mp} to classify the compared sets
      of records within a block 
      into matches and non-matches based on
      the similarity threshold $s_t$.
      The final similarities of matching sets of records 
      are sent to all the other parties $p_j$, with $2 \le j \le P$ in lines~9-10
      in Algo.~3 
      (right side of Fig.~\ref{fig-label_mpbf5}).

\begin{figure*}[!t]
  \centering
  \scalebox{2.0}[1.55]{\includegraphics[width=0.5\textwidth]
                      {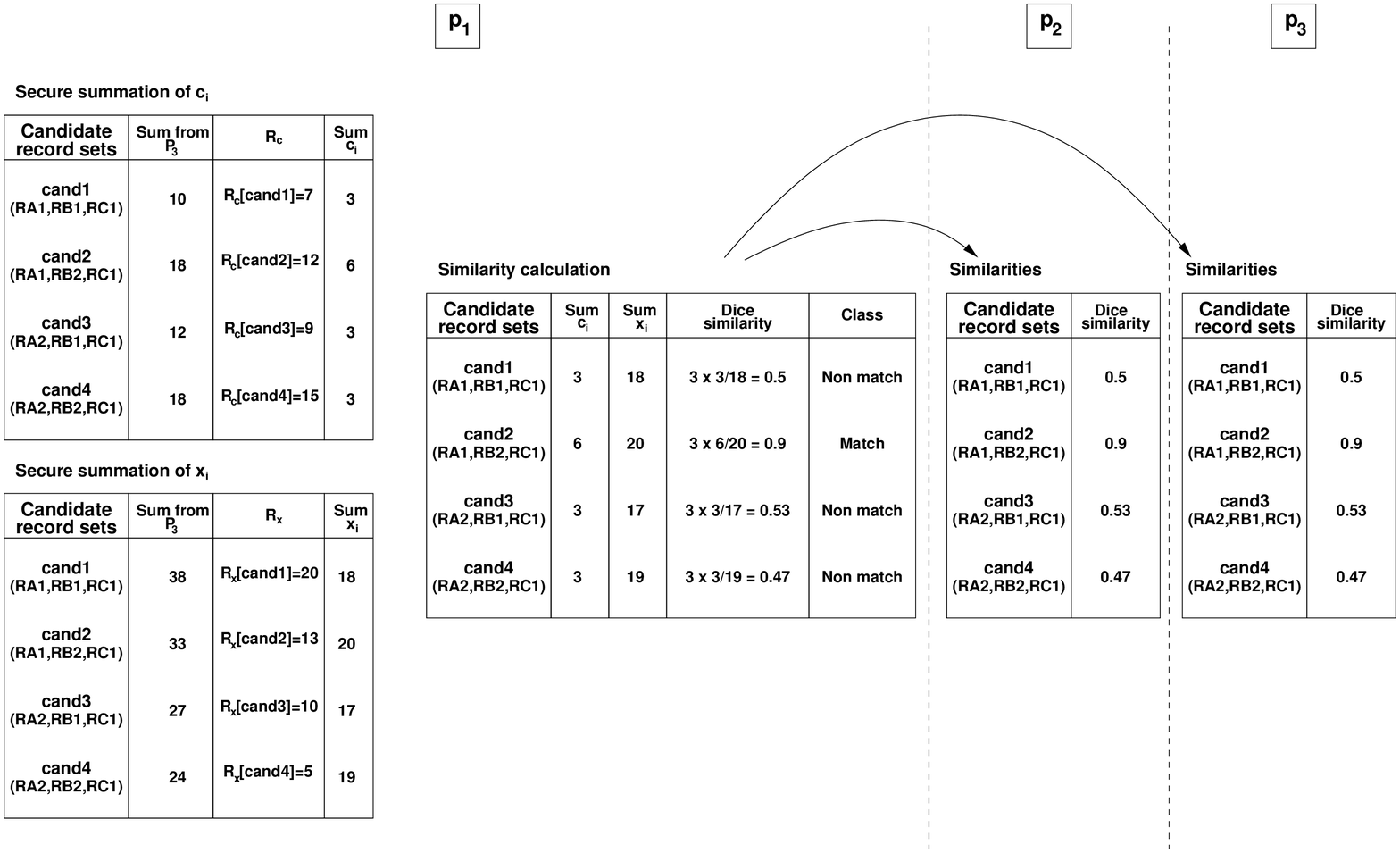}}
  \caption{The calculation of the Dice coefficient similarity of
           candidate record sets using Equation~\ref{eq:Dice_coefficient_mp}
           and the classification of sets of records into matches and non-matches.
           The minimum similarity threshold is set to $s_t=0.8$ in this example.
           We classified one matching set of records $cand2=(RA1,RB2,RC1)$ across the three datasets.
           This figure illustrates Step~7 of the protocol.
           }
           \label{fig-label_mpbf5}
\end{figure*}

\end{itemize}

\subsection{Filtering Candidate Record Sets}
\label{subsec:filter}

The most challenging aspect of multi-party PPRL is that the number of
candidate record sets can become prohibitively very large even with a
blocking technique employed. This imposes the need for using
advanced blocking and filtering approaches in order to make  
multi-party PPRL scalable and practical in real applications with
large datasets. In this section, we describe a filtering approach that
can be used in our private comparison and classification protocol to
further reduce the number of candidate record sets resulting from the
blocking step in the PPRL pipeline.

Filtering techniques are commonly employed in similarity calculations,
such as length, position and prefix filtering in PPJoin~\cite{Xia11}.
Recent work in converting such techniques into a privacy-preserving
framework~\cite{Seh15} highlighted the difficulty of applying such
traditional filtering techniques on Bloom filters. In order to achieve
high linkage quality and preserve privacy, as will be discussed in
Section~\ref{sec-analysis}, the Bloom filters used in PPRL protocols
should ideally have half of their bits set to 1 (i.e.\ be half
filled), making length, position and prefix filtering ineffective.

In our protocol, we therefore investigate the following filtering
approach which exploits the fact that parties only have access to a
fraction of all Bloom filters. 
Our assumption is that the positions of $1$-bits in the Bloom filters
are uniformly distributed 
(due to the random behavior of hash functions)
across the Bloom filters~\cite{Bloom70,Sch09}. 
This assumption of uniform distribution of $1$-bits in the Bloom filters
means that the
segments of Bloom filters of a set of records $(b_1, \cdots, b_j)$
need to have a segment similarity $seg\_sim_i(b_1, \cdots, b_j)$, with
$1 \le i,j \le P$, of at least $s_m$ in order to achieve the overall
Bloom filter similarity $Dice\_sim(b_1, \cdots, b_j) \ge s_t$ to be
classified as a matching set. 

When each party $p_i$ computes the
number of common $1$-bits, $c_i$, in the $i^{th}$ segments of each
candidate set (Step~5 of the protocol as described in
Section~\ref{sec-steps}), the party can calculate its
$seg\_sim_i(b_1,\cdots,b_P)$ as it knows the $i^{th}$ Bloom filter
segments of all $P$ records from the $P$ parties in a set. If the
$seg\_sim_i(b_1,\cdots,b_j) < s_m$ for any sub-set of $j$ ($j < P$)
records in the set of $P$ records, then the comparison and calculation
of $c_i$ can be stopped without proceeding to compare any other
sub-sets of records from the remaining $P-j$ parties with the sub-set
of records of $(b_1,\cdots,b_j)$. This basically expands lines~9-15 of
Algo.~1 as shown in Algo.~4 for party $p_i$, $1 \le i \le P$.

\begin{table}[!t]
  \label{algo_filter}

  \centering
  \small\addtolength{\tabcolsep}{-4pt}
  \begin{tabular}{ll} \hline
    ~ \\[-2mm]
    \multicolumn{2}{l}{\textbf{Algo.~4:} Filtering candidate sets by $p_i$ (extended Algo.~1).}
      \\[1mm] \hline
    ~ \\[-3mm]
    9:  & ~~ \textbf{for} $cand \in candidate\_sets$ \textbf{do}: \\ 
    10: & ~~ ~~$x_i = my\_segs.get\_x_i(cand[i])$ \\
    11: & ~~ ~~$bf\_seg = my\_segs.get\_seg(cand[i]) $ \\
    12: & ~~ ~~$seg\_x_i = bf\_seg.count\_1bits() $ \\
    13: & ~~ ~~\textbf{for} $1 \le j \le P$ and $j \neq i$ \textbf{do}: \\
    14: & ~~ ~~ ~$bf\_seg$ \& $= other\_segs.get\_seg(cand[j])$ \\
    15: & ~~ ~~ ~$seg\_c_i = bf\_seg.count\_1bits()$ \\
    16: & ~~ ~~ ~$seg\_x_i$ +$= other\_segs.get\_seg(cand[j]).count\_1bits()$ \\
    17: & ~~ ~~ ~$seg\_sim_i(b_1, \cdots , b_j) = j \times seg\_c_i / seg\_x_i$ \\
    18: & ~~ ~~ ~\textbf{if} $seg\_sim_i(b_1, \cdots , b_j) < s_m$ \textbf{then}: \\
    19: & ~~ ~~ ~~ ~ \textbf{break}\\
    20: & ~~ ~~ $c_i = bf\_seg.count\_1bits() $ \\
    21: & ~~ ~~ $\mathbf{C}_i.append([cand,c_i,x_i])$ \\[1mm]
      \hline
  \end{tabular}
\end{table}

Lines~15 to~19 show the extension of Algo.~1 for the filtering approach.
Party $p_i$ iterates over the $i^{th}$ segments of the other parties
$p_j$ ($1 \le j \le P$, $j \neq i$) in line~13.
The number of common $1$-bits and the total number of $1$-bits 
in the $i^{th}$ segments of $j$ parties are calculated in lines~14-16,
which are then used to calculate the segment similarity of $j$
segments $seg\_sim_i(b_1,\cdots,b_j)$ in line~17. If
$seg\_sim_i(b_1,\cdots,b_j) < s_m$ (line~18), then the comparison of
remaining segments ($b_{j+1}, \cdots, b_P$) with these segments
and the calculation of $c_i$ and $x_i$ can be stopped without
proceeding further, as they are with high likelihood non-matching sets.

Assuming uniform distribution of bits in the Bloom filters, $s_m$ can
be set to the same as $s_t$ so that each segment contributes the same
to the overall Bloom filter similarity $s_t$. An alternative is to set
$s_m$ to a value smaller than $s_t$ to incorporate the trade-off
between the number of false negatives (due to random hash-mapping of
$q$-grams) and the number of resulting candidate sets.

As an example of filtering, assume three databases $\mathbf{D}_1$,
$\mathbf{D}_2$, and $\mathbf{D}_3$ with records ($RA1$, $RA2$) from
$\mathbf{D}_1$, ($RB1$, $RB2$) from $\mathbf{D}_2$, and ($RC1$, $RC2$,
$RC3$) from $\mathbf{D}_3$ in the same block (resulting from a private
blocking function) need to be compared in order to identify the
matching sets of records from all three databases. This requires
private comparison of $12$ sets of Bloom filter segments from the
three databases by each party. If the $i^{th}$ Bloom filter segments
for records $RA1$ and $RB2$ do not have a similarity of at least
$s_{m}$ as calculated by party $p_i$, then the comparisons of $RA1$
and $RB2$ with $RC1$, $RC2$ and $RC3$ are not required, which reduces
the number of comparisons for a block by party $p_i$ from $12$ to $9$.
This reduction is significant when the number of parties increases
(as will be empirically shown in Section~\ref{sec-experiment}).


\section{Analysis of the Protocol}
\label{sec-analysis}

In this section we analyze our multi-party PPRL protocol
in terms of complexity, privacy, and linkage quality.

\subsection{Complexity Analysis}
\label{subsec_comp_analsis}
 
We assume $P$ parties participate in the protocol, each having a
database of $N$ records, and we assume a private blocking/indexing
technique employed in the private blocking step forms
$B \le N$ blocks for each party.
In Step~1 of our protocol, the agreement of parameters has a constant
communication complexity, and 
blocking the databases in protocol Step~2 has $O(N)$
computation complexity at each party.
Finding the intersection of blocks from all parties has a
communication complexity of $O(P\,B)$ and a computation complexity of
$O(B~log~B)$ at each party. 
Assuming the average
number of $q$-grams in the QID attributes $A$ of each record is $Q$, 
the masking of QID values of records into
Bloom filters of length $l$ using $k$ hash functions for $N$
records in Step~3 is $O(N~Q~k)$ at each party.

In Step~4, each party sends its Bloom filter
segments (each of length $l/P$) to the other parties. If we assume 
direct communication between parties, then $P(P-1)$ messages are required
in this step, each of these of size $N \times l/P$ (thus $O(N\,l\,P)$ total communication). 
With the simplified assumption that all blocks are of equal size
($N/B$),  
then in each block $(N/B)^P$ sets of Bloom filters (i.e.\ all
candidate sets of records in a block) have to be generated and their
logical conjunctions calculated in Step~5, leading to a total of $O(B(N/B)^P)$
calculations by each party. 

Filtering reduces the number of comparisons from $(N/B)^P$ to $(N/B - F)^P$,
where $F$ is the number of Bloom filter segments filtered from each party
in each block. Filtering more
non-matching record sets by increasing $F$ will improve the efficiency of
our protocol.

Steps~6 and~7 consist of the secure summation of
the calculated number of common $1$-bits ($c_i$) and total $1$-bits
($x_i$) in order to calculate the similarity of candidate sets. 
This requires for each candidate set of Bloom filters two integer numbers to be sent in
a ring communication ($P$ messages) over all parties with a total
communication of $O(P\, B(N/B - F)^P)$, followed by the distribution
of the final results which is again $O(P\, B(N/B - F)^P)$.

\subsection{Privacy Analysis}
\label{subsec_privacy_analsis}

To assess the privacy of our protocol, we assume all parties follow
the honest-but-curious adversary model~\cite{Vat13}, in that they are
curious and try to find out as much as possible about the other
parties' data while following the protocol. 
In order to analyze the privacy of our solution, 
we discuss what the parties
can learn from the data exchanged  
among them during the protocol.
There are two communication steps in our protocol
where the parties reveal some information regarding their data.

In Step~4 of our protocol, the parties split and exchange their 
Bloom filter segments (of $l/P$ length) to the corresponding other parties to
calculate the common $1$-bits in the segments.
Since calculations are
distributed among the parties, each party only learns $l/P$ bits of
each of the other parties' Bloom filters, which will make it difficult to
exploit a cryptanalysis attack~\cite{Kuz11}.
This is the highest amount of information a party can learn 
about data of other parties in our protocol.
It is important to note that
this amount of information ($1/P$ fraction of 
bits) that can be learned by a party
about another party's Bloom filters 
reduces (and thus privacy improves) with increasing $P$.

The values for the number of hash functions used ($k$) and the length
of the Bloom filter ($l$) 
provide a trade-off between the linkage quality and
privacy~\cite{Sch09}, as will be discussed in detail in
the next sub-section. The higher the value for $k/l$, the higher the
privacy and the lower the quality of linkage, because the number of
$q$-grams mapped to a single bit 
(and therefore the number of resulting collisions) increases, which leads to lower
linkage quality but makes it more difficult for an adversary to learn
the possible $q$-gram combinations~\cite{Kuz11}. 
The CLK Bloom filter encoding method
(as discussed in Section~\ref{sec-blocks}) of hash-mapping several QID
values from each record into one compound Bloom filter~\cite{Sch11,Vat12}
makes it even more difficult for an adversary to learn individual
QID values that correspond to a revealed 
bit pattern in a Bloom filter.

In addition, the parties can individually mount a simulation attack
on their own masked databases 
in the data masking and preparation step (Step~1 of our protocol)
to learn the sensitivity of each bit in their Bloom filters, as
discussed in Section~\ref{sec-steps}.
Following Durham's work~\cite{Dur13},
the sensitivity of bit position $\beta_x$, $1 \le x \le l$ in 
masked Bloom filters of $\mathbf{D}$,
is referred as $S(\beta_x)$ and calculated as:
\begin{eqnarray}
\label{eq:bit_sensitivity} \nonumber
dist(\beta_x) &=& |u|: \forall u \in \mathbf{U}~and~h_y(u) = \beta_x, 1 \le y \le k, \\ \nonumber
freq(\beta_x) &=& |r|: \forall r \in \mathbf{D}~and~r.\beta_x = 1, \\ 
S(\beta_x) &=& 1/\min \left\{dist(\beta_x),freq(\beta_x)\right\},
\end{eqnarray}
where $\mathbf{U}$ is a set of all unique $q$-grams in dataset $\mathbf{D}$,
$r.\beta_x$ is the value ($0$ or $1$) in bit position $\beta_x$ of $r$'s Bloom filter, 
and $h_y$, $1 \le y \le k$, are the hash functions used to map $q$-grams into Bloom filters.
The distribution of $q$-grams in the bits is represented by the
$dist(\beta_x)$ function which calculates the number of unique $q$-grams that are
mapped to a certain bit position $\beta_x$, and the frequency of bits
is calculated by $freq(\beta_x)$ function that counts the number of
records that set the bit position $\beta_x$ to $1$.
The minimum of these two functions is used to calculate
the sensitivity of bit $S(\beta_x)$,
since a bit that maps to a larger number of $q$-grams is not secure
(not less sensitive) if all
those $q$-grams correspond to the same record.
The higher the value for $S(\beta_x)$ is, the higher the sensitivity of bit $\beta_x$.
Based on such a sensitivity analysis, the parties can perturb their
masked datasets, for example by adding random noise~\cite{Dur13,Vat12}, to improve
the privacy of the masking at the cost of some loss in linkage quality.

The second communication step, where the
secure summation protocol is used (Step~6), requires
parties to send their sums of $c_i$ and $x_i$ values (with the
respective summed values received from the previous party) 
for each candidate set to the next party in a ring-based communication.
During this communication, however, 
no party $p_j$ can learn any information regarding
the individual values for $c_i$ and $x_i$ 
of any other party $p_i$ (with $1 \le i \le P$ and $i \neq j$),
except the final results of $\sum_{i=1}^P c_i$ and $\sum_{i=1}^P x_i$.

Party $p_1$ who initiates the secure summation protocol learns more
information than the other parties in that it can subtract the random
values and its own values from the final sums 
in order to learn $\sum_{i=2}^P c_i$ and $\sum_{i=2}^P x_i$.
Since these two results are in the range of
$0 \le \sum_{i=2}^P c_i \le (P-1)l$ and $(P-1)k \le \sum_{i=2}^P x_i \le (P-1)l$, respectively,
it would be difficult to infer the individual values of each party
due to the large number of combinations. The larger the number of parties ($P$) is,
the larger the range and the number of combinations, and thus the inference
would be harder with more parties.
The only information that $p_1$ can learn is that
if all the other segments have common
$1$-bits or not, i.e. if $\sum_{i=2}^P c_i > 0$ or not.
However, with this information, it is difficult to infer which bit positions are in common
in the other segments.

This process can also
be distributed in such a way that each party 
calculates the final sums (by initiating the secure summation protocol) for a certain
sub-set of all the candidate sets 
in order to improve the privacy of our protocol.
Another alternative approach is to use an external party to perform the secure summation
which can then send the final summed values (and the similarities) to all the $P$ parties.

\subsection{Quality Analysis}
\label{subsec_quality_analsis}

Our protocol supports approximate matching of QID values, in that
data errors and variations are taken into account depending on
the minimum similarity threshold $s_t$ used.
The quality of Bloom filter encoding based masking is dependent on the
Bloom filter parameterization~\cite{Lai06,Sch09,Dur13,Vat12}.
For a given Bloom filter length, $l$, and the number of elements $Q$
(e.g. $q$-grams) to be
inserted into the Bloom filter, the optimal number of hash
functions, $k$, that minimizes the false positive rate $f$
(of a collision of two different $q$-grams being mapped to the
same bit position), is
calculated as~\cite{Mit05}: 
\begin{equation}
\label{eqn-bf-num-hash-fun}
k = \frac{l}{Q} ln(2),
\end{equation}
leading to a false positive rate of
\begin{equation}
\label{eqn-bf-fpp}
f = \left( \frac{1}{2^{ln(2)}} \right)^{l/Q}.
\end{equation}

For a given $l$, we can calculate $k$ based on the average
number of $q$-grams, $Q$, that are generated from a record, 
as
calculated from the datasets. 
While $k$ and $l$ determine the computational aspects of our approach,
linkage quality and privacy will be determined by the false positive
rate $f$. A higher value for $f$ will mean a larger number of false
matches and thus lower linkage quality. At the same time, a higher
false positive rate $f$ will also mean improved privacy, as false
positives mean an adversary cannot be absolutely sure that a certain
bit pattern (or a Bloom filter segment) 
corresponds to a certain record~\cite{Sch09,Mit05}.

It was proven~\cite{Mit05} that a Bloom filter should ideally have
half of its bits set to $1$ (i.e.\ $50\%$ filled) to achieve the lowest
possible false positive probability for given values of $Q$, $l$ and
$k$. Equations~\ref{eqn-bf-num-hash-fun} and~\ref{eqn-bf-fpp} in fact lead
to a probability that a bit in a Bloom filter is set to $1$ as
$p = e^{-kQ/l} = 0.5$~\cite{Mit05}. For PPRL this is important,
because the bit patterns and their frequencies in a set of Bloom
filters can be exploited by a cryptanalysis attack~\cite{Kuz11}. Such
an attack exploits the fact that Bloom filters that are almost empty
can provide information about rare $q$-grams and thus rare QID
values.

In our experimental evaluation we will set the Bloom filter
parameters for our approach according to the discussion presented here
and following earlier Bloom filter work in PPRL~\cite{Sch09,Dur13,Vat12}.

\section{Experiments and Discussion}
\label{sec-experiment}

In this section, we empirically evaluate 
the performance of
our multi-party approximate matching
protocol (which we refer as `MPAM') in terms of the three properties of PPRL,
which are scalability (complexity), linkage quality, and
privacy. We use Lai et al.~\cite{Lai06}'s exact matching PPRL
approach (referred as `Lai') as a
baseline to compare with our solution, as other existing
multi-party PPRL solutions require data types
of categorical only and / or they are based on computationally
expensive SMC-based privacy techniques 
(as reviewed in Section~\ref{sec-related}).

We implemented both our proposed approach and the baseline
approach in Python 2.7.3,
and ran all experiments on a server with 
four 6-core 64-bit Intel Xeon
2.4 GHz CPUs, 128 GBytes of memory and running Ubuntu 14.04. The
programs and test datasets are available from the authors. 
Following the discussion in Section~\ref{subsec_quality_analsis} 
and other work in PPRL~\cite{Dur13,Sch09,Vat12,Vat14c}, 
we set the parameters as
$l=500$, $k=20$, $q=2$, $s_t=0.8$, and $P=[3,5,7]$. We apply a
Soundex-based phonetic blocking~\cite{Chr12} 
for the private blocking step in the PPRL pipeline
(step~1 in Fig.~\ref{fig:PPRL_multiple}),
which results in a set of blocks on which we individually conduct 
private comparison and classification (step~2 in Fig.~\ref{fig:PPRL_multiple}) 
using our approximate matching
linkage protocol.

\subsection{Datasets}
\label{sec-datasets}

To provide a realistic evaluation of our approach, we based all our
experiments on a large real-world database, the North Carolina Voter
Registration (NCVR) database as available from
\url{ftp://alt.ncsbe.gov/data/}. This database has been used for the
evaluation of various other PPRL approaches~\cite{Dur13,Vat14,Vat14b,Ran14}. We have
downloaded this database every second month since October 2011 and
built a combined temporal dataset that contains over 8 million records
of voters' names and addresses~\cite{Chr13NC}.
We are not aware of any available real-world dataset that contains
records from more than two parties that would allow us to evaluate our
multi-party approach. We therefore generated, based on the real NCVR
database, a series of sub-sets for multiple parties, 
as will be described next.

To allow the evaluation of our approach with different number of
parties, with different dataset sizes, and with data of different
quality, we used and modified a recently proposed data
corruptor~\cite{Chr13b} to generate various datasets with different
characteristics based on randomly selected records 
(with given name, surname, suburb name, and
postcode attributes as QIDs)
from the NCVR
database. 
During the corruption process we kept the identifiers of the
selected and modified records, which allows us to identify true and
false matches and therefore evaluate linkage quality of our protocol.

Specifically, we extracted sub-sets of $5,000$, $10,000$,
$50,000$, $100,000$, $500,000$, and $1,000,000$ records 
to generate datasets for $3$,
$5$, and $7$ 
parties, where
the number of matching records is set to 50\% (i.e.\ half of all
selected records occur in the datasets of all parties).
We then applied various
corruption functions in different numbers (ranging from 1 to 3) on
randomly selected attribute values which allows us to investigate how
our approximate matching
approach can deal with `dirty' data. We applied various corruption
functions, including character edit operations (insertions, deletions,
substitutions, and transpositions), and optical character recognition
and phonetic modifications based on look-up tables and corruption
rules~\cite{Chr13b}.

\begin{figure*}[ht!]
  \centering
 \includegraphics[width=0.32\textwidth]{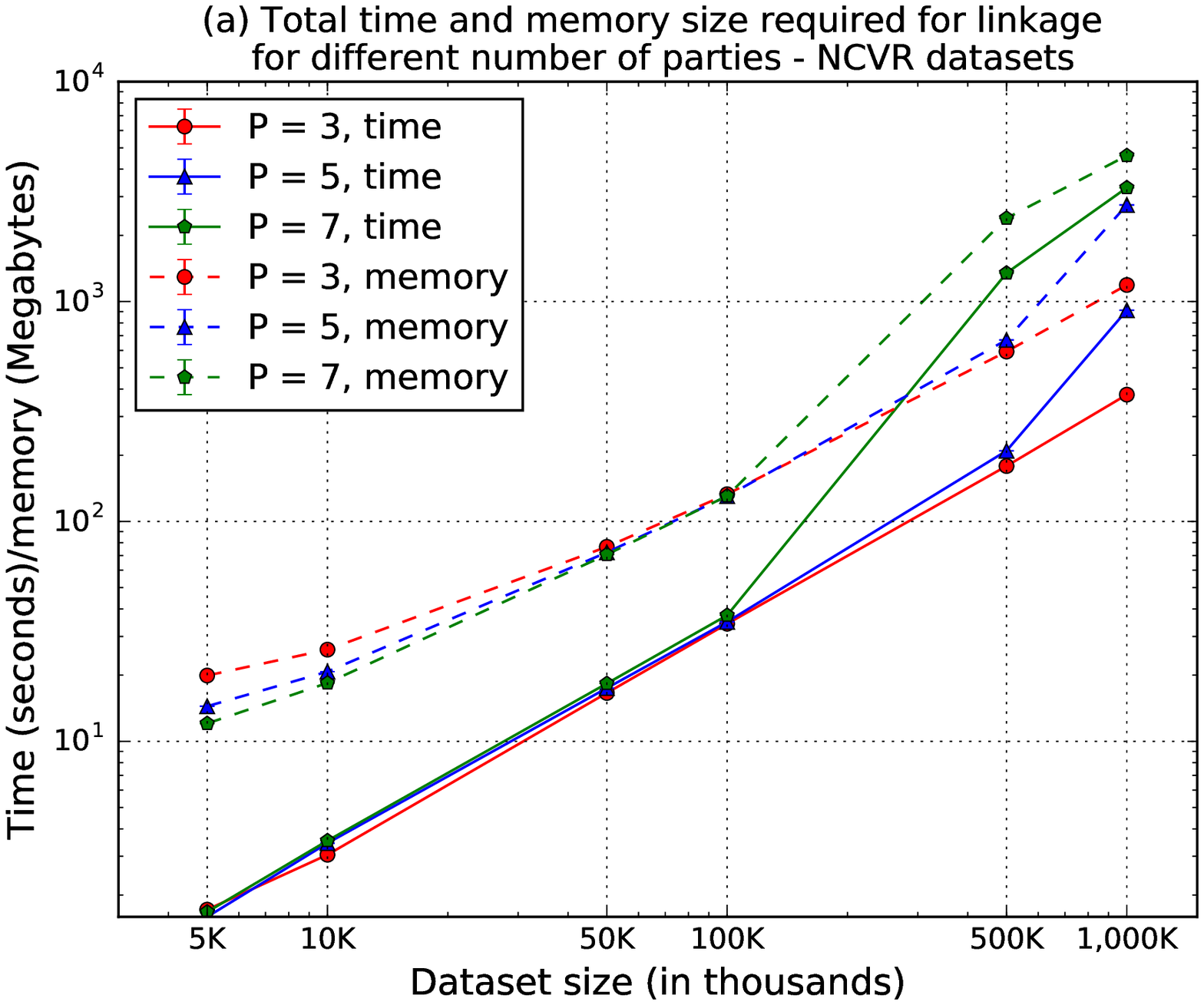}
~
 \includegraphics[width=0.32\textwidth]{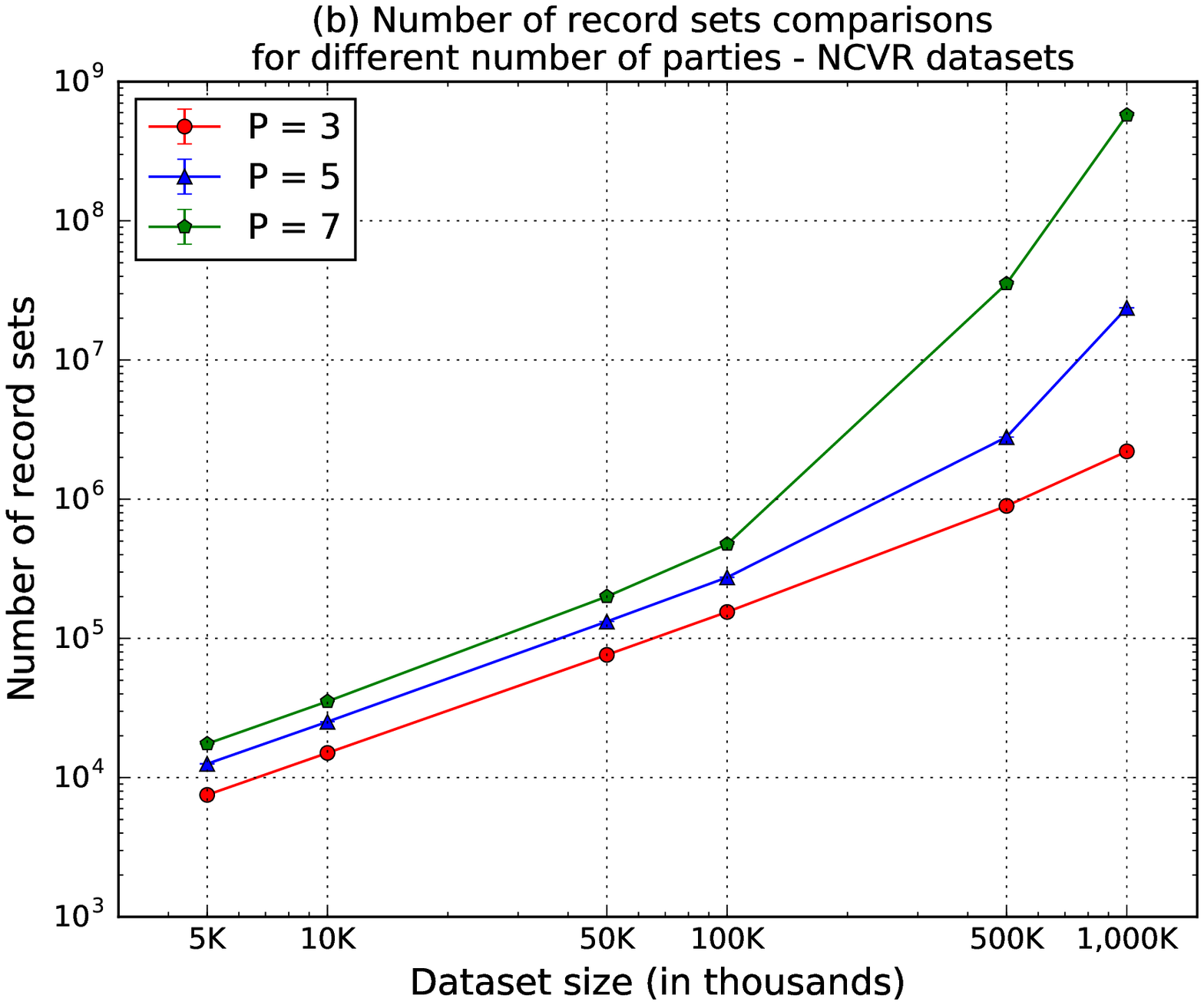}
~
 \includegraphics[width=0.32\textwidth]{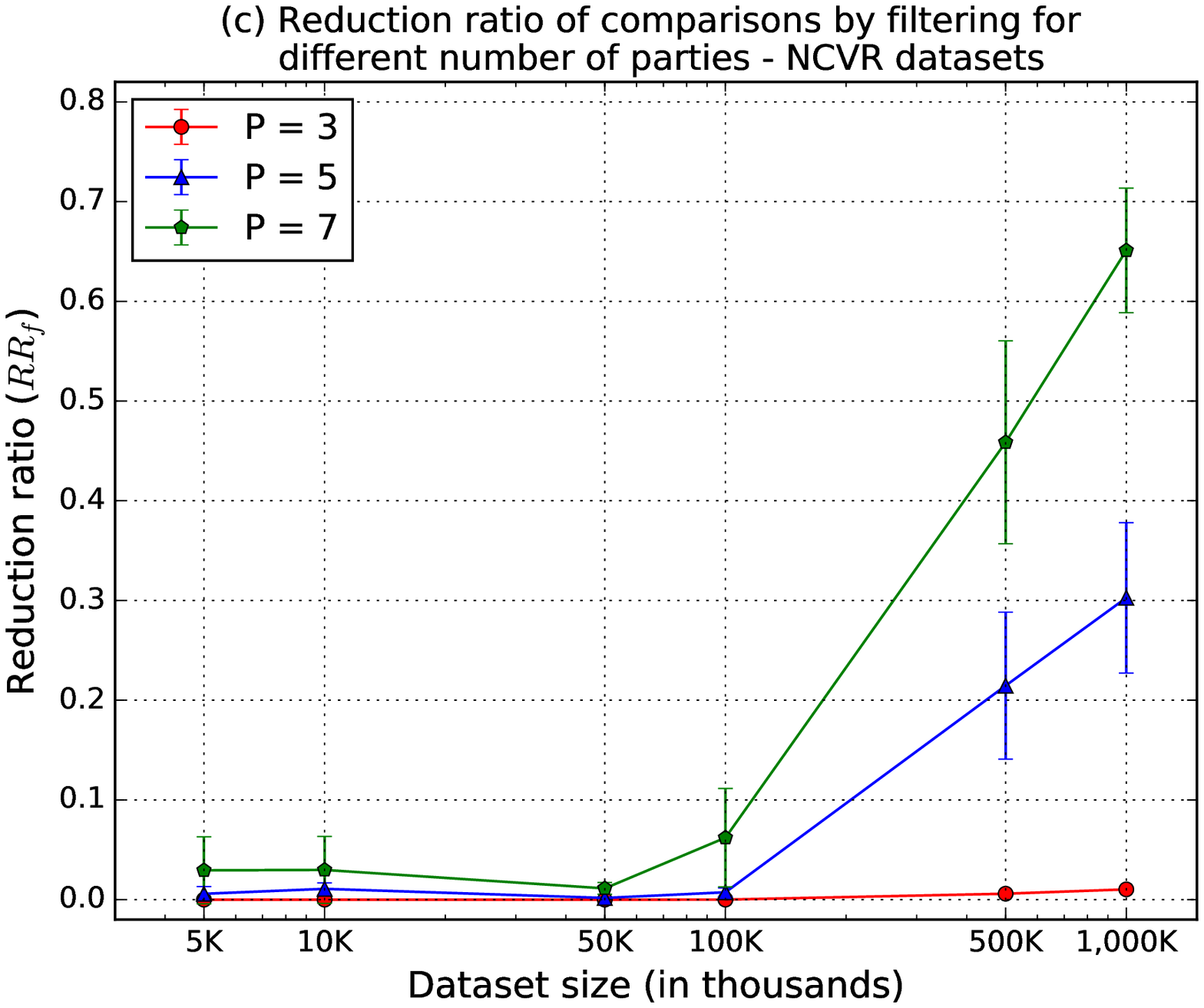}
 
  \caption{(a) Total time and memory size 
    required for linkage 
    (by one party,
    averaged over all parties), 
    (b) the number of candidate record sets
    that need to compared for the linkage, and 
    (c) the reduction
    ratio of filtered record sets ($RR_f$) by our filtering approach
    (averaged over all parties) on different dataset sizes.
    }
\label{fig:scal2}
\end{figure*}

We created several series of datasets for each of the
datasets generated above, 
where we included a varying number
of corrupted records into the sets of overlapping records ($0\%$,
$20\%$, and $40\%$). This means that a certain
percentage of records in the overlap were modified for randomly
selected parties, while the original values were kept for the
other parties. Therefore,
some of these records are exact duplicates across some parties in a
set, but are only approximately matching duplicates across the other
parties in the set. This simulates, for example, the situation where
three out of five hospitals have the correct and complete contact
details (like name and address) of a certain patient, while in the
fourth and fifth hospitals some of the details of the same patient are
different.

\subsection{Evaluation Measures}
\label{sec-measures}

We evaluate the three properties of PPRL 
for our multi-party approach using the
following evaluation measures:

The scalability of our protocol is measured by \emph{runtime} and
\emph{memory size} required for the linkage. 
Similar to the \emph{reduction ratio} ($RR$) measure
that has been used for measuring the efficiency of blocking approaches~\cite{Chr12}, 
the efficiency of our filtering approach (referred as `MPAM-F')
can be measured ($RR_f$) as follows:
\begin{eqnarray}
\label{eq:rr_f} \nonumber
RR &=& 1.0 - \frac{|candidate~sets~after~blocking|}{|all~record~sets|} \\ 
RR_f &=& 1.0 - \frac{|candidate~sets~after~filtering|}{|candidate~sets~before~filtering|} 
\end{eqnarray}

The quality of the achieved linkage is measured using the 
standard $F$-$measure$ ($F_1$) that is widely used in information
retrieval and data mining~\cite{Chr12}. 
$F$-$measure$ is the harmonic mean of $precision$ and $recall$,
calculated as~\cite{Chr12,Vat14}:
\begin{eqnarray}
\label{eq:f_meas} 
F_1 = 2 \times \frac{(precision \times recall)}{(precision + recall)},
\end{eqnarray}
where $precision$ is the fraction of
record pairs classified as matches by a decision model that are true
matches and
$recall$ is the fraction of true matches that
are correctly classified as matches by a decision model.

In line with other work in PPRL~\cite{Ran14,Vat14,Vat14b}, we
evaluate privacy using disclosure risk (DR) measures based on the
probability of suspicion ($p_s$), i.e.\ the likelihood a masked database
record in $\mathbf{D}^M$ can be matched with one or several (masked) record(s) in a
publicly available global database $\mathbf{G}$. 
The probability of suspicion for a masked value/record $r^M$, $p_s(r^M)$,
is calculated as $1/n_g$ where
$n_g$ is the number of possible matches in $\mathbf{G}^M$
to the masked value $r^M$.
We conduct a frequency linkage attack~\cite{Vat14} on our protocol
using equivalent datasets as used in the linkage $\mathbf{D}$
to be the global databases
(i.e. $\mathbf{G} \equiv \mathbf{D}$ in the worst case, because when
$\mathbf{G} \equiv \mathbf{D}$ there will be one-to-one exact matching
of global value for each value in $\mathbf{D}$)
by mapping the revealed bit patterns (segments) in the Bloom filters
in $\mathbf{D}^M$
to the Bloom filters in $\mathbf{G}^M$
in order to calculate the
following disclosure risk measures, as proposed by Vatsalan et al.~\cite{Vat14}.

\begin{itemize}

\item \emph{Mean disclosure risk} ($DR_{Mean}$): This takes into consideration the
      distribution of probability of suspicion of all values in $\mathbf{D}^M$ and is calculated as
      the average risk ($\sum(p_s)/|\mathbf{D}^M|$) of any sensitive value being re-identified.

\item \emph{Marketer disclosure risk} ($DR_{Mark}$): This is calculated as the proportion of masked records 
(Bloom filter segments) in $\mathbf{D}^M$ that
match to exactly one masked record in $\mathbf{G}^M$ ($|\{r^M \in \mathbf{D}^M: p_s(r^M) = 1.0\}| /|\mathbf{D}^M|$).
\end{itemize}


\subsection{Experimental Results}
\label{sec-results}

Figs.~\ref{fig:scal2} (a) and~\ref{fig:scal2}(b) 
show the scalability of our
approach, measured by runtime and memory size
required for the linkage
as averaged over all parties, and
the number of candidate record sets to be compared
and classified for the linkage. 
Runtime slightly increases with larger number of parties ($P$) and is
almost linear in the size of the datasets.
Interestingly, memory size decreases with
$P$ because the Bloom filter segments at
each party become shorter ($l/P$) and the similarity calculations are
distributed among the parties. 
However, memory size increases on the larger datasets with larger $P$,
because the number of record sets becomes large
with more parties even with the 
phonetic blocking~\cite{Chr12} and our filtering approach
(as described in Section~\ref{subsec:filter})
employed. 

\begin{figure*}[ht!]
  \centering
 \includegraphics[width=0.32\textwidth]{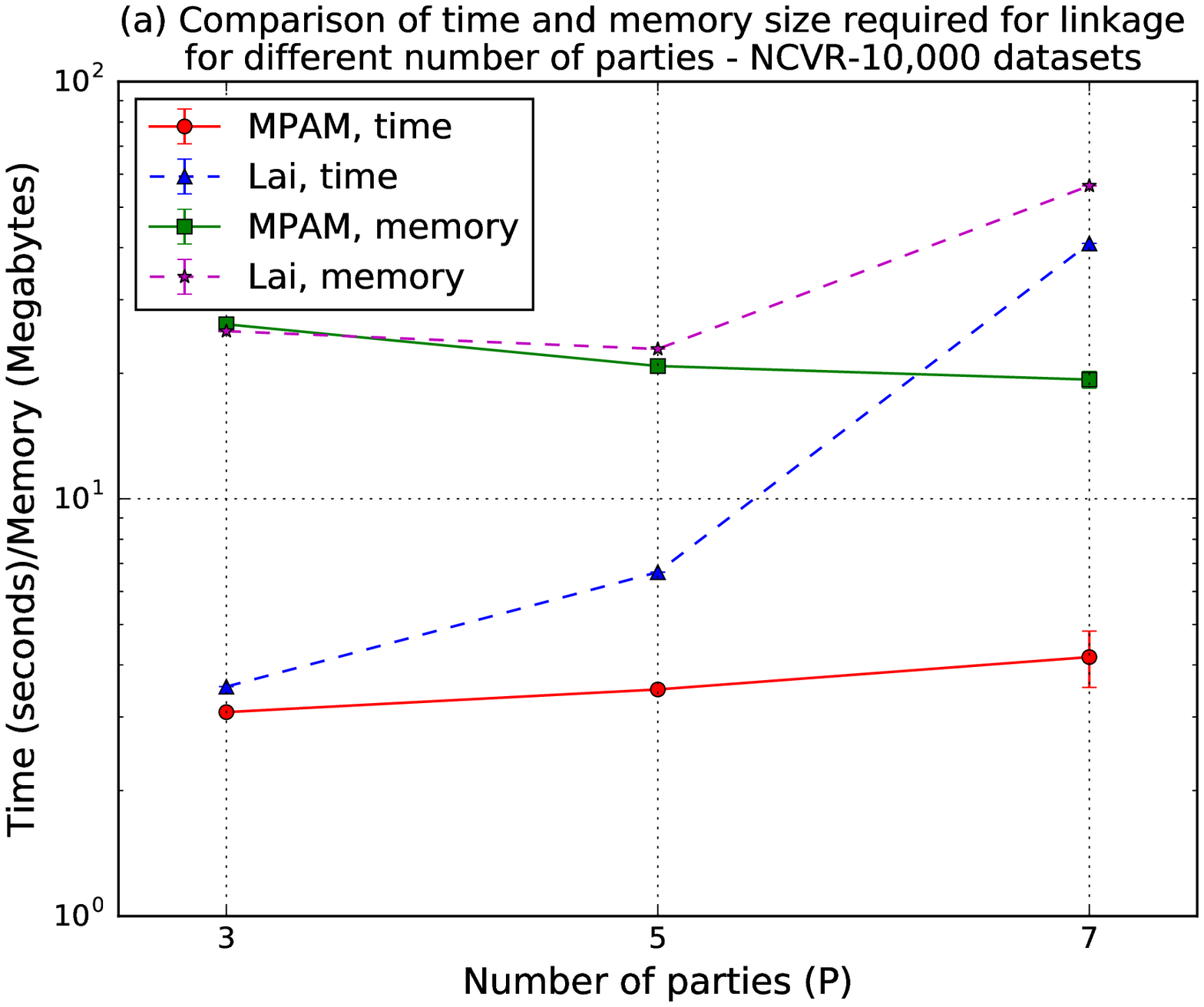}
~
 \includegraphics[width=0.32\textwidth]{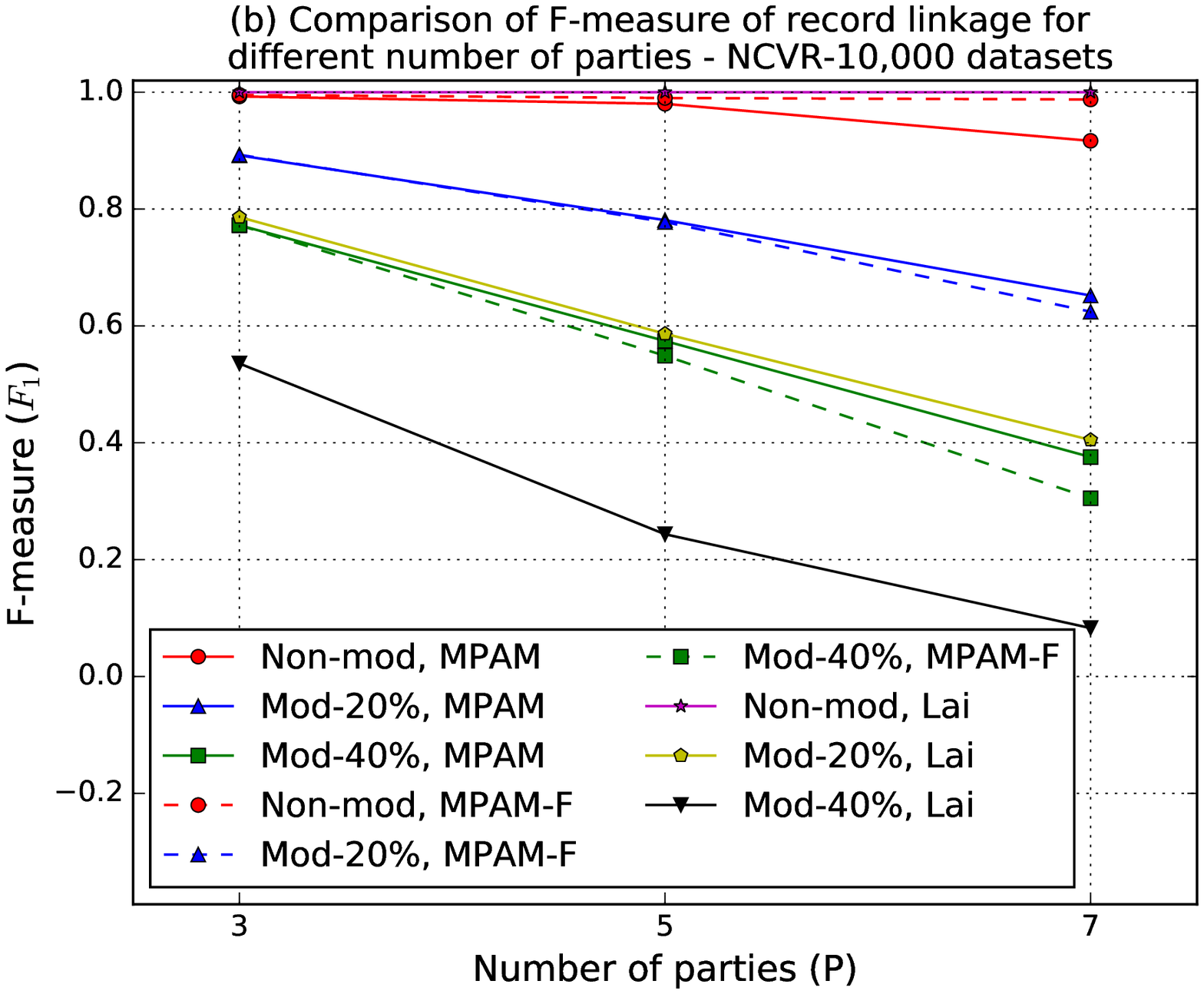}
~
 \includegraphics[width=0.32\textwidth]{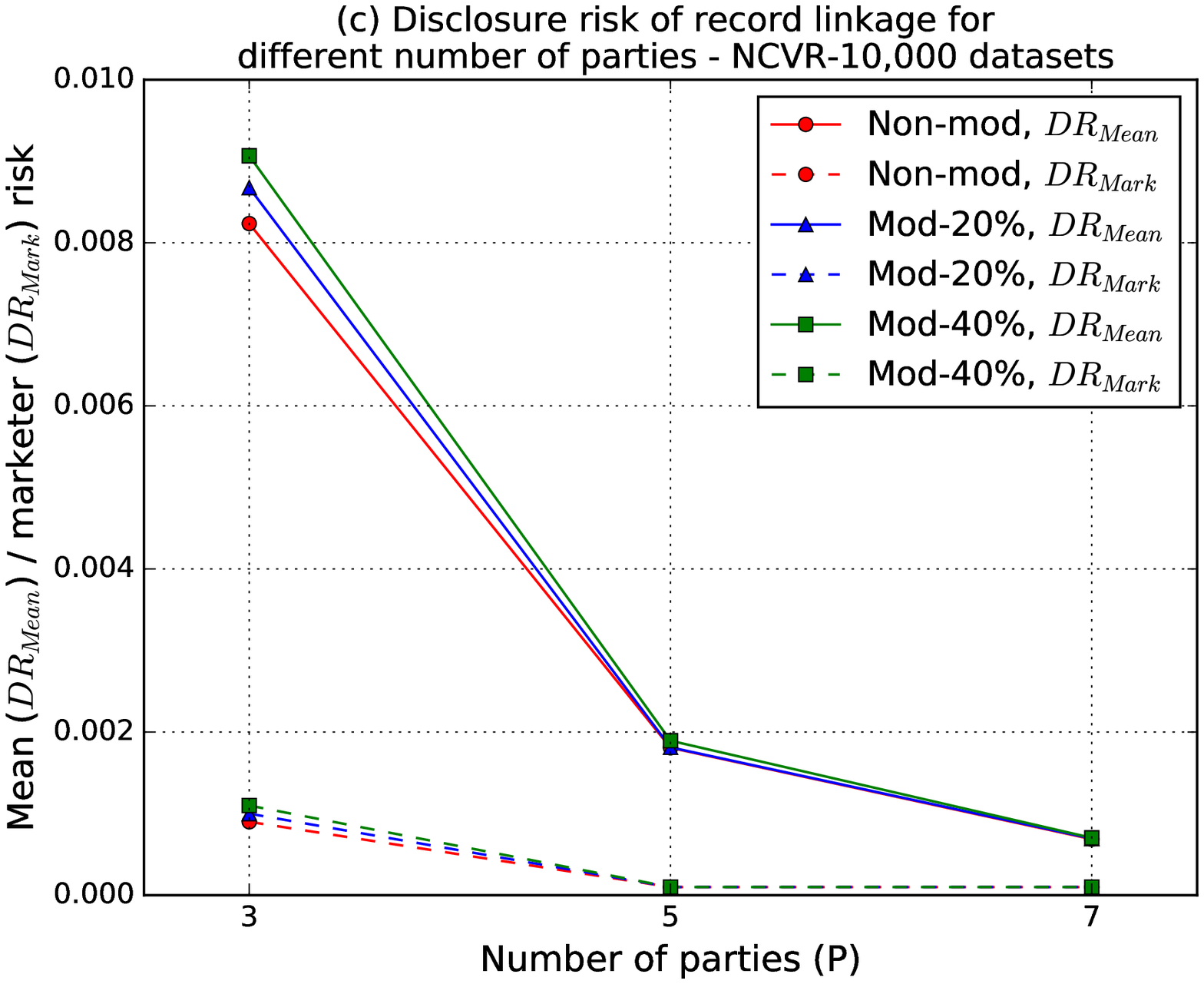}
 
  \caption{Comparison of (a) runtime and memory size required, 
     (b) F-measure achieved, and (c) disclosure risk measures~\cite{Vat14}
     of privacy, with our proposed approach (MPAM), 
     combined with filtering approach (MPAM-F), and Lai et al.'s approach~\cite{Lai06} (Lai), 
     on the NC-10,000 datasets with all variations
     for different number of parties.
    }
\label{fig:accuracy}
\end{figure*}

The reduction ratio of record set comparisons ($RR_f$) by
our filtering approach 
is shown in Fig.~\ref{fig:scal2} (b).
The $RR_f$ is not significant on smaller datasets. However,
it achieves a moderate $RR_f$
in the number of comparisons on larger datasets and it
increases with the number of parties, $P$.
This opens up a research direction of developing advanced
filtering and blocking/indexing approaches, and
efficient communication patterns
for multi-party PPRL techniques
to be studied further.

Compared to the baseline approach by Lai et al.~\cite{Lai06}, 
our approach is more
scalable and efficient in terms of linkage time
and memory size, and is linear in the number of parties,
as shown in Fig.~\ref{fig:accuracy} (a). The reason is that
in the baseline approach by Lai et al.,
after the Bloom filter segments are distributively
processed by the parties 
to compute the segments with common $1$-bits (conjuncted),
each party has to perform a membership test of its own
Bloom filters with the conjuncted Bloom filters in order
to classify them as matches or non-matches~\cite{Lai06}.
However, in our proposed approach only one party 
(or alternatively an external party) calculates
the similarities of record sets based on the sums 
of the number of $1$-bits and common $1$-bits of all
parties, which are then distributed to all parties.
An interesting aspect is that the time required by our approach
slightly increases and memory size
decreases with more parties 
(increasing $P$),
while they
increase significantly with the baseline approach.

The quality of linkage, as measured using the F-measure ($F_1$),
achieved with our approach
(both MPAM and MPAM-F) and the baseline approach (Lai)
is compared
in Fig.~\ref{fig:accuracy} (b) on the NCVR-10,000 datasets. 
As can be seen from the figure, $F_1$ is high
on the non-modified datasets ($0\%$ corruption). 
On the modified datasets (with $20\%$ and $40\%$ corruption) $F_1$
drops quite drastically with the number of parties. The reason is that 
when records with modifications occur in each dataset the number of missed true
matching record sets increases. 
The filtering approach (MPAM-F) only affects
the quality of the linkage slightly, as we achieved similar results to
MPAM.
On the non-modified datasets
the filtering approach performs comparatively well. This is because
the precision improves by removing false matching sets, and thus leads
to higher $F_1$ results.
Though the baseline approach performs well on the non-modified
datasets, $F_1$ is significantly lower on the modified
datasets (as the baseline approach by Lai et al. supports only exact matching). 

Finally, the privacy of our protocol (as well as Lai et al.'s approach~\cite{Lai06}), 
as measured by DR
measures~\cite{Vat14}
(mean disclosure risk and marketer disclosure risk),
for a frequency linkage attack on Bloom filter segments
in the NCVR-10,000 datasets
to the known values in the global database $\mathbf{G}$
(in the worst case setting of $\mathbf{G} \equiv$ NCVR-10,000) for different
number of parties is shown in
Fig.~\ref{fig:accuracy} (c). 
As discussed in
Section~\ref{subsec_privacy_analsis}, 
disclosure risk decreases (i.e.\ privacy
increases) with an increasing number of parties $P$ 
as the Bloom filter segments ($l/P$) become shorter and are therefore
matched to a larger number of global records (i.e. $n_g$ increases). 
This results in lower probability of suspicion of segments ($p_s = 1/n_g$) 
with larger $n_g$ and provides lower
values for the disclosure risk measures~\cite{Vat14}.


\section{Conclusions and Future Work}
\label{sec-conclusion}

We have presented an efficient and approximate private comparison and
classification protocol for multi-party PPRL 
based on Bloom filter encoding and distributed
secure summation. Our protocol efficiently
identifies sets of records that have 
a high Dice coefficient similarity across all 
the parties. The protocol has a communication complexity that is
linear in the number of parties and the size of the databases that are
linked, making the protocol scalable to 
applications where data from multiple parties need to be linked.
However, a main bottleneck of multi-party PPRL is the large number of
candidate record sets.

In future work, we plan to improve the scalability of our protocol by 
reducing the number of candidate record sets further 
using improved and advanced private blocking or filtering approaches, and by
investigating different communication patterns.
A second avenue of future work will be to conduct 
linkage attacks
on the protocol with different Bloom filter encoding methods
and different noise addition techniques to further evaluate the
privacy of our approach. 
In terms of linkage quality, we also
plan to investigate how to make our protocol more general and allow
for different approximate string similarity functions~\cite{Chr12}
to be incorporated.
Developing PPRL techniques
for identifying matching record sets across sub-sets of multiple
databases is another important research direction.
Finally, we plan to investigate improved
classification techniques for multi-party PPRL
including relational clustering and
graph-based approaches~\cite{Chr12} which are successfully used in
non-PPRL applications.

Our ultimate aim is to develop techniques that allow for large
databases to be linked in secure, accurate, automatic, and scalable
ways across many parties, thereby facilitating novel ways of data
analysis and mining that currently are not feasible due to privacy and
confidentiality concerns.



\ifCLASSOPTIONcompsoc
  \section*{Acknowledgments}
\else
  \section*{Acknowledgment}
\fi

This research is funded by the Australian Research Council Discovery Project DP130101801.

\ifCLASSOPTIONcaptionsoff
  \newpage
\fi

\bibliographystyle{IEEEtran}
\bibliography{paper}

\end{document}